\documentclass[prl,reprint,showpacs,superscriptaddress,amsmath,amssymb]{revtex4-1}
\usepackage{epsfig}
\usepackage[]{natbib}
\usepackage{xspace}
\usepackage{color}
\usepackage{xcolor}
\usepackage{bbold}
\usepackage{tabu}
\usepackage{mathtools}
\usepackage[colorlinks=true,
	    urlcolor=blue,
            linkcolor=blue,
            anchorcolor=blue,
            citecolor=blue
            ]{hyperref}

\DeclarePairedDelimiter\bra{\langle}{|}
\DeclarePairedDelimiter\ket{|}{\rangle}
\DeclarePairedDelimiter\bracket{\langle}{\rangle}

\begin{document}

\title{Restart expedites quantum walk hitting times}

\author{Ruoyu Yin} 
\affiliation{Department of Physics, Institute of Nanotechnology and Advanced Materials, Bar-Ilan University, Ramat-Gan
52900, Israel}
\author{Eli Barkai}
\affiliation{Department of Physics, Institute of Nanotechnology and Advanced Materials, Bar-Ilan University, Ramat-Gan
52900, Israel}

\begin{abstract}

Classical first-passage times under restart are used 
in a wide variety of models, 
yet the quantum version of the problem still misses key concepts. 
We study the quantum hitting time with restart
using a monitored quantum walk.  
The restart strategy eliminates the problem of dark states, 
i.e. cases where the particle evades detection, 
while maintaining the ballistic propagation 
which is important for fast search.
We find profound effects of quantum oscillations on the restart problem, 
namely a type of instability of the mean detection time,
and optimal restart times that form staircases, 
with sudden drops as the rate of sampling is modified. 
In the absence of restart and in the Zeno limit, 
the detection of the walker is not possible 
and we examine how restart overcomes this well-known problem,
showing that the optimal restart time 
becomes insensitive to the sampling period. 

\end{abstract}
\maketitle

{\em Introduction.}
First-passage processes are ubiquitous in practically all fields of science. 
Probably the simplest approach uses a random walker in search for a target, 
as found for diffusion-controlled reactions \cite{Redner2001,Ralf2014book}. 
This common method does not demand an input of energy 
however it is non-efficient as the random walker 
resamples previously-visited locations  
and further the walker according to the laws of chance 
may stray far from the target. 
In the context of biochemical reactions
nature found a way to overcome this problem, 
and that is with a restart strategy 
\cite{Bel2009a,Bel2009,pnas,MMOptRestart}.  
It turns out that sometimes, if the search does not find its target
it is better to give up, and start the process anew.
Restarts were employed to accelerate algorithms \cite{Luby1993,Gomes1998},
and then considered in generality in the context of stochastic processes 
\cite{Majumdar2011,Evans2011},
rapidly encompassing various contexts including classical search theory, 
chemical physics and population dynamics, etc. 
In this well-studied field the basic questions are
what are the non-equilibrium steady states emerging from restart,
and what is the optimal time to restart 
\cite{Luby1993,Gomes1998,Majumdar2011,Evans2011,Evans2014,Denis2014,Gupta2014,Eule2016,
Reuveni2016,restart,Majumdar2017,Belan2018,Igor2018,Evans2018,Boyer2019,
Igor2019a,Igor2019b,Lukasz2019,Majumdar2020,Majumdar2020a,Evans2020,
Redner2020,Tal2020,Eliazar2021,Redner2021,Daniel2021,Majumdar2021,Ralf2021a,Ralf2021}.

\begin{figure}[htbp]{}
\centering
\includegraphics[width=0.9\columnwidth]{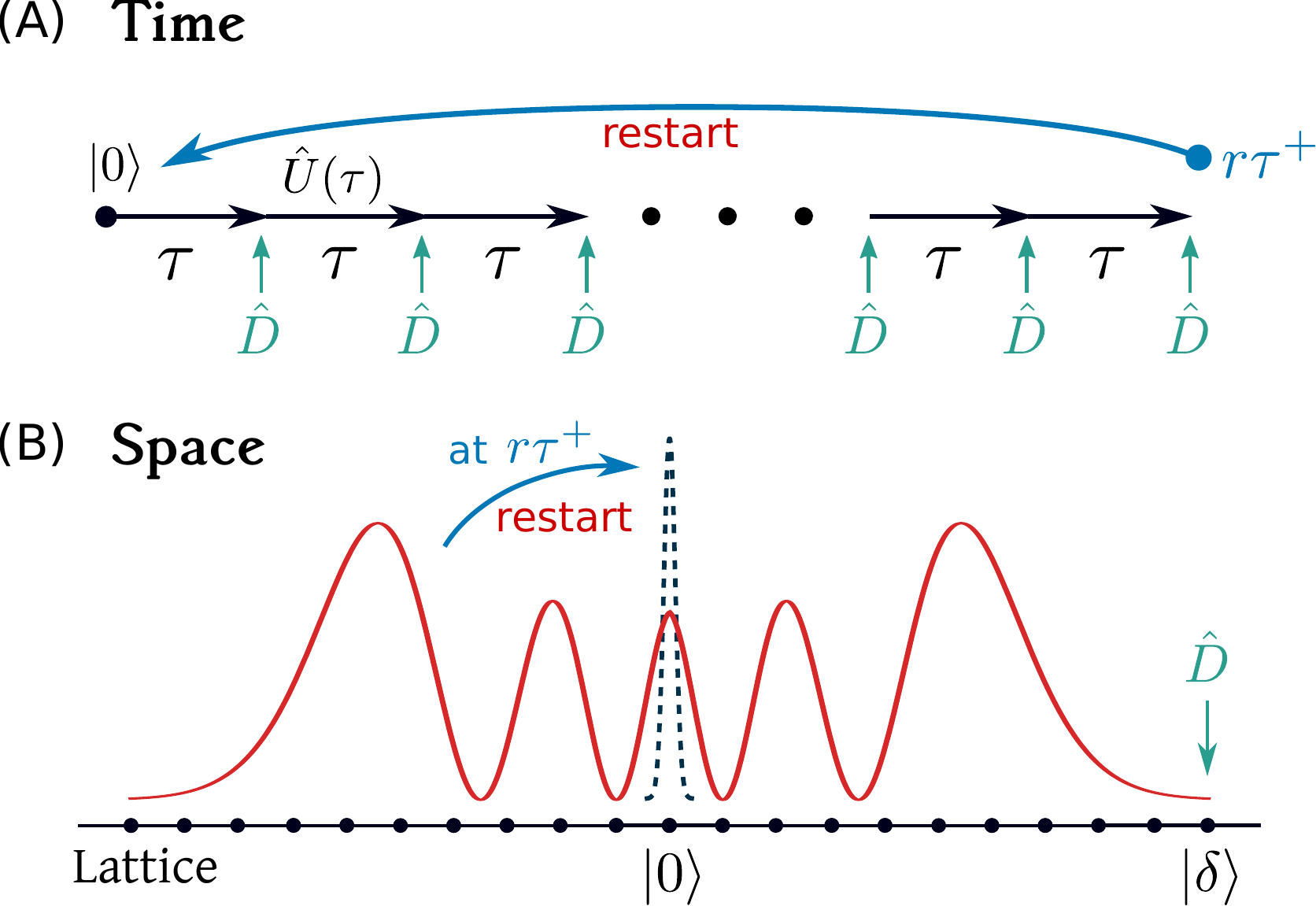}
\caption{The measurement protocol under sharp restart 
in time (A) and space (B) representation.
The system is initialized at state $\ket{0}$ 
[see also the dashed needle in (B)]
and the unitary evolution $\hat{U}(\tau)$ 
is repeatedly interrupted by projective measurements defined with
$\hat{D}=\ket{\delta}\bra{\delta}$ at times
$\tau,2\tau,3\tau,\cdots$. 
If the state is detected we are done, 
if not at the $r$-th failure of detection, 
the system is
brought back to $\ket{0}$, 
i.e. a restart is performed every $r$ steps.
The red curve in (B) represents the wave packet, 
namely the solution of Eq. (\ref{eq2}), 
that spreads out ballistically before the particle is detected or reset.
}
\label{fig0}
\end{figure}
As the counterpart of classical random walks, 
quantum walks are widely applied in many different fields, 
ranging from transport in waveguides to ultra-cold atoms 
to light-harvesting dynamics in biochemistry
\cite{Aharonov1993,Farhi1998,Kempe2003,Yaron2008,Mulken2011},
%
and therefore rather naturally a few previous works addressed 
the restart problem with an underlying quantum dynamics 
\cite{Majumdar2018,Rose2018,Belan2020,Perfetto2021,
Perfetto2021a,Turkeshi2021,Magoni2022}. 
At the same time, quantum search and transport, in the absence of restarts, 
has an antagonist: the dark subspace \cite{Caruso2009,Thiel2020D}, 
caused by destructive interference.
This problem works against the quantum advantage 
of ballistic propagation \cite{Yaron2008}, which can be useful for search.
This means that for quantum walks with dark states, the detection probability, defined below,  
is less than unity even for small systems.
Overcoming this hurdle is important for efficient quantum search 
and restarts are a powerful approach for that aim.
However, one may not introduce restarts blindly, 
as the goal is not merely to get rid of the dark states, 
but rather to optimize the time for search. 
The basic questions are: 
How to choose the time for the restart 
so that the quantum search time is minimized?
Will the ballistic superiority of the quantum search be retained when restart is added? 
What are the fundamental differences between quantum and classical restarts?
To characterize the time for search, we utilize the concept of quantum hitting time,
or the first-detected-passage time (FDPt).
The model we consider is a tight-binding quantum walk with repeated monitoring,
which was studied extensively in the absence of restart 
\cite{Krovi2006,Gruenbaum2013,Dhar2015,Dhar2015a,Friedman2017a,Thiel2018a,
Thiel2020J,Thiel2020D,Dhar2021,Das2022}. 
Such repeated monitoring/measurements 
have been implemented for example on IBM quantum computers \cite{Sabine2022}.
Our work paves the way to speedup of quantum hitting times, on quantum computers, 
which as mentioned is particularly important in the presence of dark states.

\begin{figure}[htbp]{}
\centering
\includegraphics[width=\columnwidth]{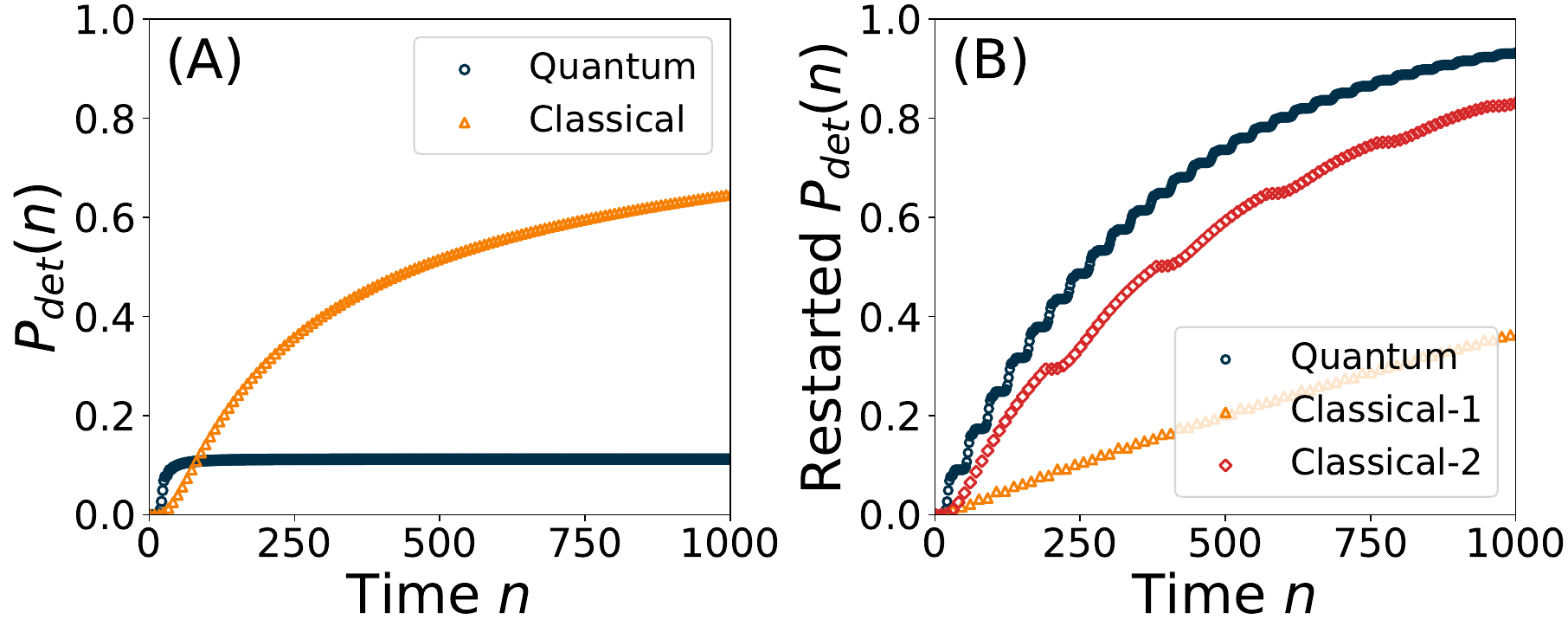}
\caption{(A) 
Detection probability $P_\text{det}(n)$ 
for a classical/quantum walker on an infinite 1D lattice.
The quantum total detection probability
$P_\text{det} = \sum_{n=1}^\infty F_n \approx 0.1$, 
although it grows rapidly at the beginning.
The figure illustrated that for short times, 
the quantum ballistic spreading speeds up the search 
(compared to the classical counterpart) 
but at long times the quantum detection without restart performs poorly. 
(B) Restarted quantum walk performs by far better 
than the corresponding classical walk,
when $r=35$ for both models (classical-1).
The quantum restart process also performs better 
if compared with the optimally chosen classical restart 
(classical-2 when $r=191$). 
Here $\tau=0.25$, $\delta=10$.
}
\label{fig1}
\end{figure}
The benchmark model for classical restart is a diffusive particle 
whose position is reset at some random time $t_r$
\cite{Majumdar2011,Evans2011,Igor2018}. 
In this case the mean time the particle reaches a fixed target is finite 
(without restart it diverges in an unbounded domain).
Further, the mean time to reach a target has a distinct minimum, 
and thus optimal value, as a function of $t_r$
\cite{Majumdar2011,Evans2011,Evans2014,restart,Gupta2022}. 
In contrast, using quantum walks,
we will find several minima instead of a unique minimum, 
and the mean FDPt can exhibit a bi-stable behavior 
where the hitting time is optimal for a pair of values of $t_r$.

Consider a classical random walk on the integers, and let $P^{{\rm  cl}}(x,t)$
be the probability of finding the particle on $x$ at time $t$. 
The master equation is \cite{Redner2001}
\begin{equation}
{\partial_t P^{{\rm cl }} (x,t) 
} = 
{\gamma  } \left[P^{\rm cl}(x+1,t)
+  P^{\rm cl}(x-1,t)
- {2} P^{\rm cl}(x,t) \right].
\end{equation}
$\gamma$ is the hopping rate.
Now consider the tight-binding quantum walk, 
the wave-function
is $\ket{\psi(t)}= \sum_{-\infty} ^\infty \phi(x,t) \ket{x}$ and 
using the Schr\"odinger equation,
\begin{equation}\label{eq2}
-i {\partial_t \phi(x,t) 
} = 
\gamma \left[ \phi(x+1,t) + \phi(x-1,t) - 2 \phi(x,t) \right]
\end{equation}
with $\hbar=1$.
In both models the walker hops to nearest neighbors 
and the solutions for starting from the origin $\ket{x}=\ket{0}$ 
are $\phi(x,t) = i^x e^{-i2\gamma t} J_{x}(2\gamma t)$ 
\cite{Friedman2017a},
$P^{{\rm cl}}(x,t) = i^{-x} e^{-2\gamma t} 
J_{x}(i 2\gamma t)$ \cite{Redner2001},
where $J_\alpha(z)$ is the Bessel function of the first kind. 
Replacing $t$ with $i t$ 
one may switch from $P^{{\rm cl}}(x,t)$ to $\phi(x,t)$. 
Still, the packet spreadings are different: 
the classical packet spreads diffusively 
and approaches a Gaussian for large times,  
while the quantum walk propagates ballistically \cite{Yaron2008}.

Quantum systems generally lack precise trajectories. 
Hence to define first-hitting time we add repeated monitoring
at $\ket{\delta}$,
with the goal to detect the particle at this state. 
For that an observer makes repeated measurements at times 
$\tau, 2 \tau, \cdots$.  
Each measurement is a projection, 
namely either the particle is found at $\ket{\delta}$ (yes) 
or it is not (no), see Fig. \ref{fig0}. 
This yields a string of measurements, no, no, $\cdots$
and in the $n$-th attempt a yes \cite{Krovi2006,Dhar2015,Friedman2017a}. 
The process of search is then completed. 
Clearly the first time we get a click yes is random,
and $n\tau$ is defined as the hitting time or the FDPt.
Note that classically the continuous sampling 
of the process $\tau\to 0$ makes sense,
but with the quantum framework this leads to a freeze of the dynamics 
and to null detection due to the quantum Zeno effect \cite{Misra1977}.

Let $F_n$ be the probability of detecting the walker 
in the $n$-th attempt for the first time without restart. 
Classical and quantum renewal equations were extensively used
to obtain these basic probabilities 
\cite{Redner2001,Friedman2017a}. 
The quantum $F_n$'s are presented in Table \ref{tab2}, 
see also details below and the SM.
To start we plot in Fig. \ref{fig1}A 
the detection probability up to time $n\tau$, i.e. 
$P_{{\rm det} }(n) = \sum_{n^\prime=1} ^n F_{n^\prime}$ still without restart
(with $\gamma=1$).
We see that at short times $n\tau$,
the quantum walker is performing better, 
as it has the advantage of ballistic propagation. 
However, at large times the classical walker wins 
in the sense that it is eventually detected 
with probability one while the quantum system falls far from this limit 
\cite{Thiel2018a}.  

To improve the hitting time, we use the sharp-restart strategy
\cite{Luby1993,restart}, leaving other cases to SM, see also further discussion at the end of the letter.
Every $r$ detection attempts we restart the search process,
as Fig. \ref{fig0} depicts. 
With this approach we find both simple and novel results, we start with the former. 
Using the simple example in Fig. \ref{fig1}A, 
if we choose $r$ to be slightly larger than the time 
it takes the quantum $P_{{\rm det}} (n)$ to saturate without restart, 
we observe two effects presented in Fig. \ref{fig1}B. 
First, the quantum detection is now guaranteed: 
with probability one we detect the walker in the long time limit 
{(the same for the classical cases)}. 
Second, the quantum walker performs much better than the classical one (classical-1), 
in the sense of a much larger quantum $P_{{\rm det}} (n)$ 
compared with the classical case. 
If $r$ is chosen as the optimal for the classical walk 
to make a fair comparison, 
the quantum restart process still performs better than the classical one 
(classical-2).
This is obviously due to the quantum ballistic propagation.

To gain insight, we will focus on the expected FDPt 
under restart, denoted by $\langle t_f\rangle_r$.
By definition $\langle t_f\rangle_r = \tau\langle n_f\rangle_r
=\tau(r \langle {\cal R} \rangle + \langle \tilde{n} \rangle)$,
where $n_f$ is the number of measurements until first hitting,
${\cal R}$ is the number of restarts before final detection.
Hence $0\le{\cal R}\le\infty$ and $1\le\tilde{n}\le r$.
The joint distribution of ${\cal R}$ and $\tilde{n}$ is 
${\rm Pr}_{r}({\cal R},\tilde{n})=[1-P_{\rm det}(r)]^\mathcal{R} F_{\tilde{n}}$,
with the normalization 
$\sum_{{\cal R}=0}^\infty \sum_{\tilde{n}=1}^r {\rm Pr}_{r}({\cal R},\tilde{n})=1$. 
Using the restart time $t_r=r\tau$, 
we obtain \cite{Luby1993,Eliazar2021,Pal2021} 
\begin{equation}\label{maineq}
\langle t_f \rangle_r
=
\underbrace{{ 1-P_{\rm det}(r) \over P_{\rm det} (r)} t_r \vphantom{\sum_{\tilde{n}=1}^r n} }_{\langle{\cal R} \rangle t_r}
+ 
\underbrace{\sum_{\tilde{n}=1}^r {(\tilde{n}\tau) F_{\tilde{n}} \over P_{\rm det} (r)} }
_{\tau \langle \tilde{n} \rangle}.
\end{equation}
(see SM for detailed derivation).
In turn, the probabilities $F_n$ were studied previously 
in Refs. \cite{Friedman2017a,Thiel2018a} and as mentioned
for small $n$ are presented in Table \ref{tab2} (for $\delta=0$).
The $F_n$'s are used to evaluate observables of interest numerically,
though clearly as a stand-alone quantity they do not provide much insight. 
We now focus on the optimization for $\langle t_f \rangle_r$ 
in small and large $\tau$ limits 
where applicable approximations allow analytical solutions.

\begin{table}
\centering
\caption{$F_n$ for the model of an infinite line, $\delta=0$.}
\begin{tabular}{ll}
\hline \hline $\mathrm{n}$ & \multicolumn{1}{c}{$F_{n}$} \\
\hline 1 & $|J_{0}(2 \gamma \tau)|^{2}$ \\
2 & $|J_{0}(4 \gamma \tau)-J^{2}_{0}(2 \gamma \tau)|^{2}$ \\
3 & $|J^{3}_{0}(2 \gamma \tau)-2 J_{0}(4 \gamma \tau) J_{0}(2 \gamma \tau)+J_{0}(6 \gamma \tau)|^{2}$ \\
4 & $|-J^{4}_{0}(2 \gamma \tau)+3 J_{0}(4 \gamma \tau) J^{2}_{0}(2 \gamma \tau)
-2 J_{0}(6 \gamma \tau) J_{0}(2 \gamma \tau)$ \\
& $-J^{2}_{0}(4 \gamma \tau)+J_{0}(8 \gamma \tau)|^{2}$ \\
\hline\hline
\end{tabular}
\label{tab2}
\end{table}
\begin{figure}[htbp]{}
\centering
\includegraphics[width=0.91\columnwidth]{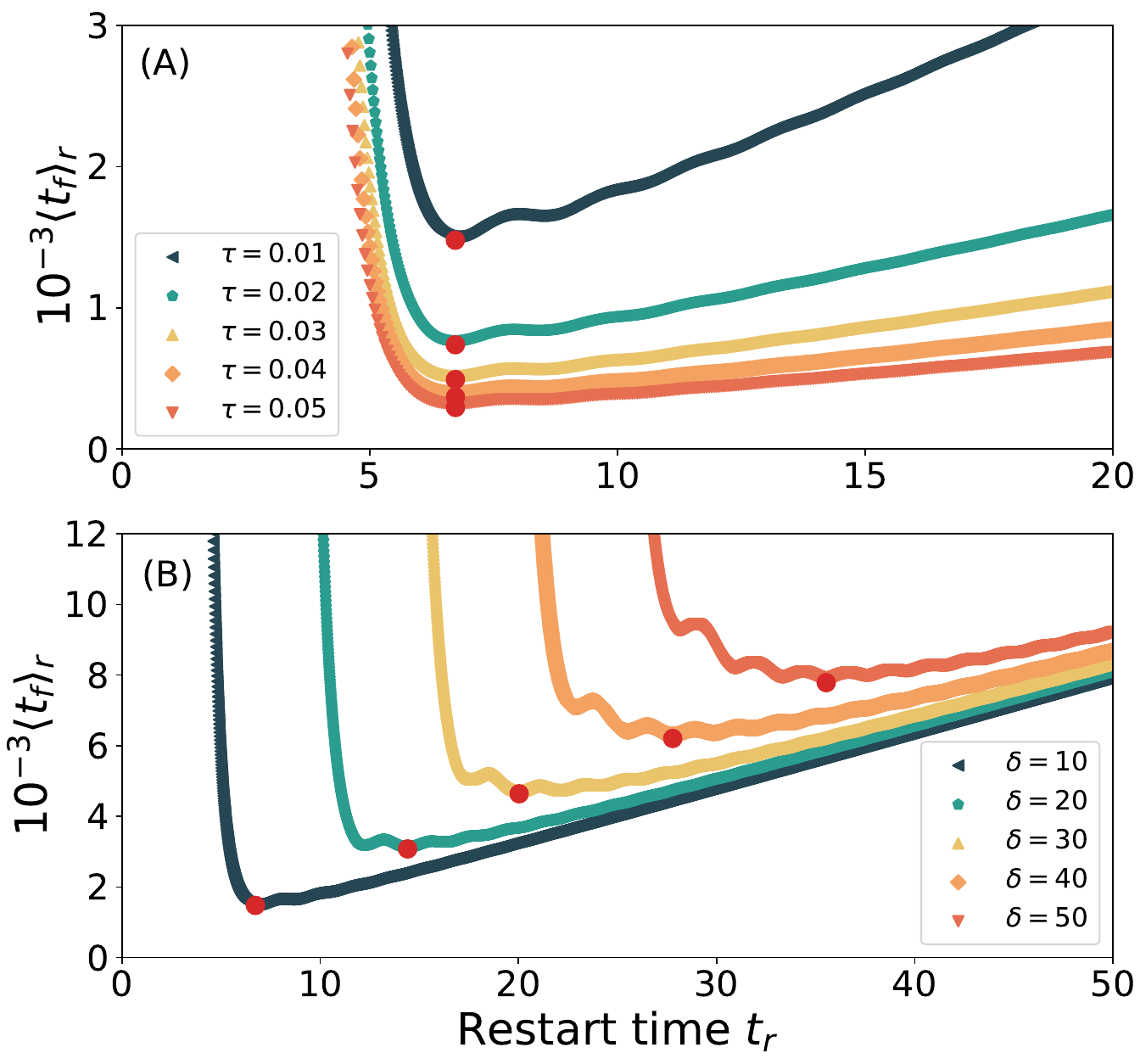}
\caption{$\langle t_f \rangle_r$ vs. restart time $t_r$ 
for different $\delta$ and $\tau$,
obtained using the repeated-measurement model. 
The red dots are the minimum calculated 
using the non-Hermitian approximation Eq. (\ref{eq6}).  
Notice the oscillations which are a quantum feature.
We used $\delta=10$ (A), 
and $\tau=0.01$ (B).
The optimal restart time is $\tau$ independent (A) 
while exhibiting a nearly linear dependence on the distance $\delta$ (B). 
}
\label{fig2}
\end{figure}
{\em Zeno limit.} 
We now consider the case where $\tau$ is small 
and hence the measurements are frequent. 
Let $g(t_f) {\rm d} t_f$ be the probability of $t_f$, 
in the absence of restart, 
to be in the interval $[ t_f, t_f + {\rm d}t_f]$.  
In this limit, the process can be modeled with a continuous time formalism, 
which is a great simplification \cite{Dhar2015,Dhar2021}. 
This means that we can treat  
$g(t_f)= \sum_{n=1}  F_n \delta(t_f - n \tau)$ 
as a smooth function.
The main tool here is a non-Hermitian Schr\"odinger equation, 
$i \dot{\ket{\Psi}} = [H - i\hbar(2/\tau)\ket{\delta}\bra{\delta}]\ket{\Psi}$,
and $S(t)= \bra{\Psi(t)} \Psi(t)\rangle$ 
is the probability that the walker ``survived'' from detection until time $t$, 
and $g(t_f)= - {\rm d} S(t) / {\rm d} t|_{t=t_f}$. 
With this approach and $H$ defined in Eq. (\ref{eq2}) 
one finds that without restart 
\cite{Thiel2020J,Krapivsky2014} 
\begin{equation}
g(t_f) = \tau \, \delta^2 {J^2_\delta(2t_f)\over t_f^2}.
\label{eqg1}
\end{equation}
As $\tau\to0$,
this expression exhibits the well-known Zeno physics,
i.e. frequent measurement prohibits state transitions \cite{Misra1977}. 
To remedy this problem we use restart.

Using Eq. (\ref{maineq}) in the continuous limit yields
\begin{equation}\label{eq6}
\begin{aligned}
      \langle t_f \rangle_r \sim&  
      \left[
      \left(
      \delta^2\tau {\cal I}_1
      \right)^{-1}-1 
      \right]
      t_r 
      + 
      {\cal I}_2/{\cal I}_1 
      ,
  \end{aligned}
\end{equation} 
where ${\cal I}_1 =\int_0^{t_r} {t_f^{-2} J^2_\delta(2t_f)}{\rm d}t_f$,
${\cal I}_2 = \int_0^{t_r} {t_f^{-1} J^2_\delta(2t_f)}{\rm d}t_f$
(see SM for explicit solutions).
The theory Eq. (\ref{eq6}) nicely matches numerical results
obtained from the repeated-measurement model (see SM). 
In Fig. \ref{fig2}, we present $\bracket{t_f}_r$ as a function of $t_r$, 
on top of which the global minimum of $\langle t_f \rangle_r$ (red dots) is provided 
via minimization of Eq. (\ref{eq6}),
which remarkably is $\tau$ ($\delta$) independent (dependent) respectively.
We will soon explain this intriguing feature. 

We also see in Fig. \ref{fig2} that for too small or too large $t_r$, 
the mean hitting time diverges as expected.
Specifically, using Eq. (\ref{eq6}), for $\delta\neq0$,
$\bracket{t_f}_r \sim (2\delta-1)\Gamma^2(1+\delta)  
t_r^{2(1-\delta)}/\tau\delta^2$
when $t_r\to0$, 
and $\bracket{t_f}_r \sim \left[{\pi(4\delta^2-1) /8\tau\delta^2}-1\right]t_r
+ {(4\delta^2-1)\pi / 16\delta}$ as $t_r\to\infty$.
Hence for large $t_r$ and large $\delta$, 
$\bracket{t_f}_r$ is linear in $t_r$
with a $\delta$-independent slope,
which is vastly different from the classical behavior,
as the latter is proportional to $\delta$ \cite{Majumdar2011}.

The detection time under restart features oscillations, 
clearly visible in Fig. \ref{fig2}. 
These oscillations are in turn related to the phase acquired in 
$g(t_f) \sim (\delta^2 \tau /\pi)t_f^{-3}\cos^2(2t_f-\pi\delta/2-\pi/4)$
in large $t_f$ limit of Eq. (\ref{eqg1}).
The quantum oscillations presented in Fig. \ref{fig2} imply that we have  
in general multiple extrema for $\bracket{t_f}_r$, 
instead of a unique minimum usually found for classical restarts 
\cite{Majumdar2011,Evans2011,Evans2014,restart,Gupta2022}, 
see also SM where we present classical examples. 
Using Eq. (\ref{eq6}) the extrema are solutions to
\begin{equation}
\begin{aligned}
\tau \xi - \tau^2 
	\left[ 
        \xi \int_0^{t_r^{ext}} \tilde{g}(t){\rm d}t 
        +
        \int_0^{t_r^{ext}} t\, \tilde{g}(t){\rm d}t 
    \right]
=0
,
\end{aligned}
\end{equation}
where $\tilde{g}(t) = g(t)/\tau$,
$\xi= \int_0^{t_r^{ext}} \tilde{g}(t){\rm d}t 
- t_r^{ext} \, \tilde{g}(t_r^{ext})$,
and the superscript $ext$ means extremum.
Since  $\tau$ is small we may neglect the $\tau^2$ terms 
and find a transcendental equation for the extrema, i.e. $\xi=0$ or
\begin{equation}\label{cont1}
\int_0^{t_r^{ext}} {J^2_\delta(2t)\over t^2} {\rm d}t 
=  {J^2_\delta(2t_r^{ext})\over t_r^{ext}},
\end{equation}
(see SM for an explicit solution to the integral).
Hence the extrema are independent of $\tau$, 
as demonstrated in Fig. \ref{fig2}. 
%
Note that similar technique to derive the optimum is used also from classical walks 
\cite[Eq. (7)]{Igor2018}.
Remarkably, since as mentioned after Eq. (\ref{eq2}) 
$|J_\delta(2t)|^2$ is the probability of finding the walker at $\ket{\delta}$,
{\em in the absence of measurements}, 
the extremal restart times are actually connected to the solution of 
the Schr\"odinger equation $|\phi(\delta,t)|^2$. 
The transcendental Eq. (\ref{cont1}) indicates that  
the number of extrema increases as $\delta$ grows. 
Unlike the classical problem, the global minimum $t_r^*$ increases 
roughly linearly with the distance $\delta$,
and exhibits sudden jumps at special $\delta$'s 
due to the multiple minima (see SM).

\begin{figure}[htbp]{}
\centering
\includegraphics[width=\columnwidth]{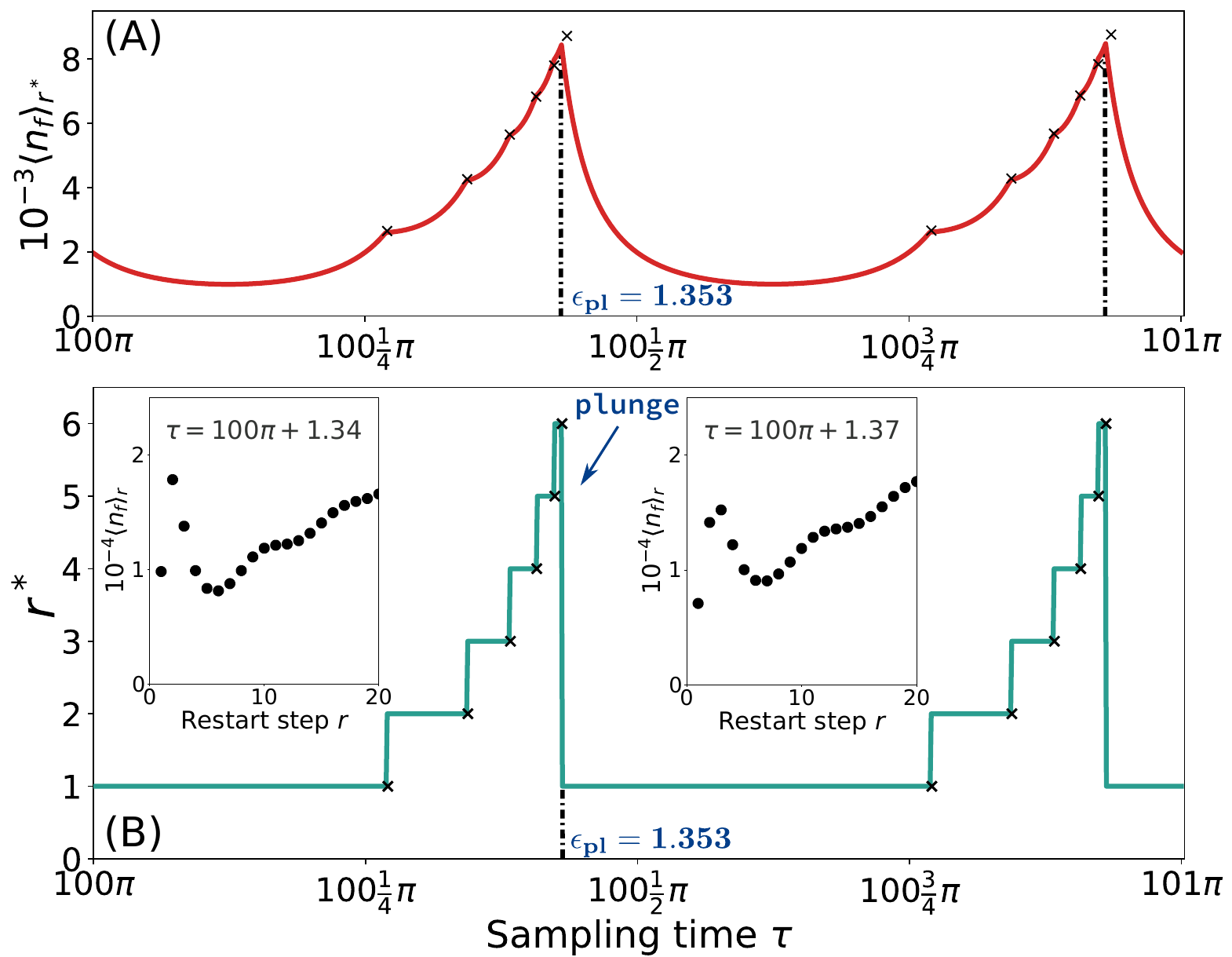}
\caption{(A) $\langle n_f \rangle_{r^{*}}$ versus $\tau$.
$\langle n_f \rangle_{r^{*}} \sim 1/ F_{r^*+1}$ is used 
to calculate the theoretical optima at transition points
(black crosses),
around which nonsmoothness is witnessed.
(B) The optimal restart step $r^*$ versus $\tau$.
We see the novel staircase structure in large $\tau$, 
however this type of behavior appears in the full range of $\tau$ (see SM).
The black crosses represent the theoretical transition $\tau$'s Eq. (\ref{eq13}).
For classical restart, $r^*=1$.
The insets present $\langle n_f \rangle_r$ vs. $r$ 
in the vicinity of the plunge $\tau$ ($=100\pi+1.353$). 
There are two minima competing with each other, 
and a small change of $\tau$ (i.e. $\Delta \tau=0.03$) results in different optima.
On the left inset $r^*=6$ and on the right $r^*= 1$.
We used $\delta=0$.
}
\label{fig4}
\end{figure}

{\em Large $\tau$ limit.} 
Also when the measurement period $\tau$ is large 
we find interesting effects. 
In this case and without restart 
the probability of FDPt $F_n$
is given by the wave-function of the system in the absence of measurements. 
The origin of this effect is that sparse measurements do not modify 
the Hermitian dynamics too much. 
Specifically, using the asymptotics of the Bessel function 
\cite{Friedman2017a,Remark}
\begin{equation}
F_n (\tau) = |\bra{0} \psi(t=n \tau) \rangle|^2 
\sim {1 \over n \pi \tau} 
\cos^2 \left( 2n\tau - {\pi \over 4} \right).
\label{eqFnbig}
\end{equation}
Here we have focused on the case called the return problem when $\delta=0$, 
partly due to space limitation.  
In Fig. \ref{fig4} we plot $r^*=t_r^\ast/\tau$ versus $\tau$
using a numerically exact calculation.
Clearly unlike the Zeno case, now $\tau$ is an important parameter. 
Remarkably as shown in Fig. \ref{fig4}, 
$r^*$ exhibits a periodic sequence of staircases, which are now analyzed.

Beyond the fact we get for the optimum $r^*$ periodic-like behavior, 
there are plunges for certain critical $\tau$'s in Fig. \ref{fig4}. 
This means that the optimal restart step jumps from $r^*=6$ to $r^*=1$, 
when $\tau$ is only slightly modified.
This indicates the existence of instabilities in the system, 
related to quantum oscillations.
To understand these effects, we used Eq. (\ref{eqFnbig}).
First notice that choosing $2\tau$ as a multiple of $\pi$, 
we have $F_n \sim 1/ (2 n \pi \tau)$. 
Since $F_n$ is monotonically decaying with $n$, it is not difficult to realize
that the optimal strategy to restart is to choose $r^*=1$, 
namely immediate restart 
(this holds in the classical counterpart 
since the first-passage probability decays as $t^{-3/2}$ \cite{Redner2001}).
This explains the periodicity presented in Fig. \ref{fig4}. 
As we vary $\tau$, then close to $\tau = k \pi/2$ with $k$ a positive integer, 
the best strategy is to restart as fast as possible,
i.e. $r^*=1$.

What will happen when we increase $\tau$? 
Considering the mean $\langle n_f \rangle_r= \langle t_f \rangle_r/\tau$,
note that $\langle n_f \rangle_1= 1/F_1$ 
and importantly when $\tau$ is large, 
such that $F_n\ll1$ we have
$\langle n_f \rangle_2 \sim 2/ (F_1 + F_2)$, 
$\langle n_f \rangle_3 \sim 3/ (F_1 + F_2 + F_3)$ etc. 
Let $\tau=k \pi/2 + \epsilon$ and $0<\epsilon<\pi/2$. 
As mentioned for $\epsilon=0$,   
$r^*=1$.
When the condition  $\langle n_f \rangle_1= \langle n_f \rangle_2$ holds,
there is a transition from $r^* = 1$ to $r^*=2$ 
taking place when $\epsilon_{1 \to 2} =0.850$
(see derivation in SM). 
Further transitions in $r^*$ from $r$ to $r+1$ take place whenever 
\begin{equation}
r F_{r+1} = \sum_{n=1} ^r F_n.
\label{eqRRR}
\end{equation}
%
Importantly this formula admits a finite number of solutions, 
which based on physical intuition is expected, 
since $r^*$ cannot be too large.
Using Eqs. (\ref{eqFnbig},\ref{eqRRR})  
we find those $\epsilon_{j\to j+1}$
on which the step-like jumps take place, 
with $j$ increasing from $1$ till $5$,  
\begin{equation}\label{eq13}
	\{\epsilon_{j\to j+1}\} = \{0.850, 1.081, 1.204, 1.280, 1.332 \}, 
	\epsilon_{pl}= 1.353.
\end{equation}
The subscript ${pl}$ means plunge, i.e. the jump from $r^*=6$ to $r^*=1$. 
Thus as shown in Fig. \ref{fig4} we have a complete theory 
of the staircase structure.
Further using the smallness of $F_n$ we have at transition points
$\langle n_f \rangle_{r^{*}} \sim 1/ F_{r^*+1}$.

To better understand the ``plunge''
namely the transition $r^*= 6\to r^*= 1$ found for $\epsilon_{pl}$,
$\langle n_f \rangle_r$ close to a critical value of $\tau$ is plotted
in the insets of Fig. \ref{fig4}. 
There appear two minima of $\langle n_f \rangle_r$. 
The first is at $r^*=6$ and the second at $r^*=1$. 
A slight change of $\tau$ leads to
the global minimum switching from one value to the other. 
At the exact transition value,
the two minima are identical. 
Thus the system exhibits an instability in the sense that 
small changes of $\tau$ create a large difference in $r^*$.

{\em Discussion.}
Employing the sharp-restart strategy, 
we expedite the hitting time of a tight-binding quantum walk
and now we emphasize three points.
First, the expected hitting time under restart exhibits an oscillatory behavior 
unlike the classical case, rendering the appearance of several extrema. 
This effect is general and not limited to the Zeno limit, 
as we will show in a future publication.
What is unique to this limit, 
is that the optimal restart is $\tau$-independent, 
and that one may obtain a transcendental equation 
which is a by far simpler tool 
if compared with an exact though numerical evaluation of the problem.
Second,
previously it was shown that sharp restart has certain advantage 
of attaining the lowest mean passage time among all restart strategies
\cite{Luby1993,restart}.
It is also noteworthy that the quantum feature of oscillations 
is wiped out with Poisson or geometric restarts  
(see details in SM),
thus sharp restart should be used in the quantum domain.
Third,
in sparse measurement limit, i.e. large $\tau$,
the optimal restart step $r^*$ exhibits a periodical staircase structure 
with instabilities, i.e. plunges in the optimal restart time 
(Fig. \ref{fig4}). 
We expect these instabilities to be generic 
for a wide range of parameter changes, 
as their cause is the oscillatory nature of the detection time statistics.
These plunges and instabilities are clearly a signature of the quantum dynamics, 
and as far as we know, are new in the general framework of restart theory. 

Our theory can be implemented in laboratories, 
as restarts are routinely used, 
for the aim of repeating experimental protocols
to gain statistics of various outputs. 
Probably the best way to test the theory is on quantum computers. 
Here the repeated strong measurements needed for hitting time statistics and the restarts 
i.e. the returning of the system to its initial state, 
are now built in parts of quantum computing package. 
The quantum walk part is implemented by the Jordan-Wigner transformation, 
that maps the walk to a qubit representation \cite{Sabine2022}. 
It should be noted that system size does not have to be large, 
as some of the effects we found here, 
like staircases and plunges (Fig. \ref{fig4}) are generic to all quantum systems.

\begin{acknowledgments}
The support of Israel Science Foundation's grant 1614/21 is acknowledged.
\end{acknowledgments}


%

\appendix 

\section{General stochastic restart strategies}
\begin{figure}[htbp]{}
\centering
\includegraphics[width=0.88\linewidth]{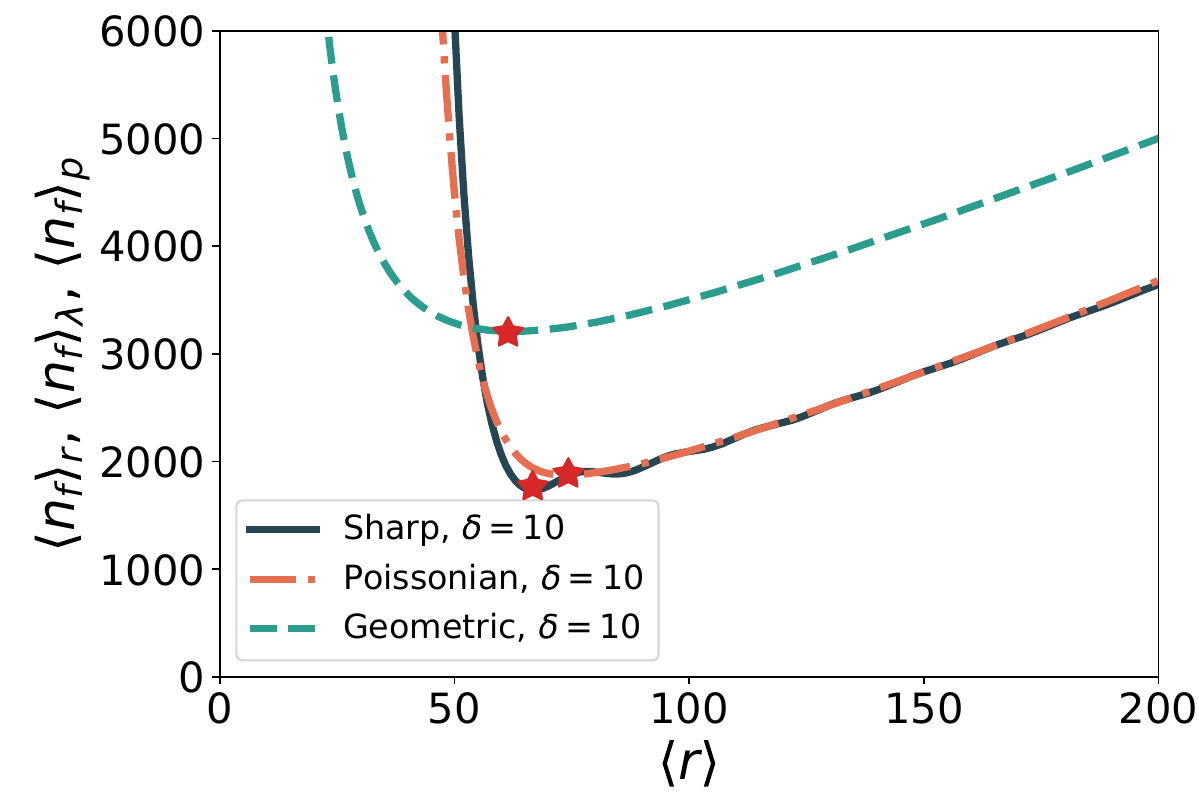}\\
\includegraphics[width=0.88\columnwidth]{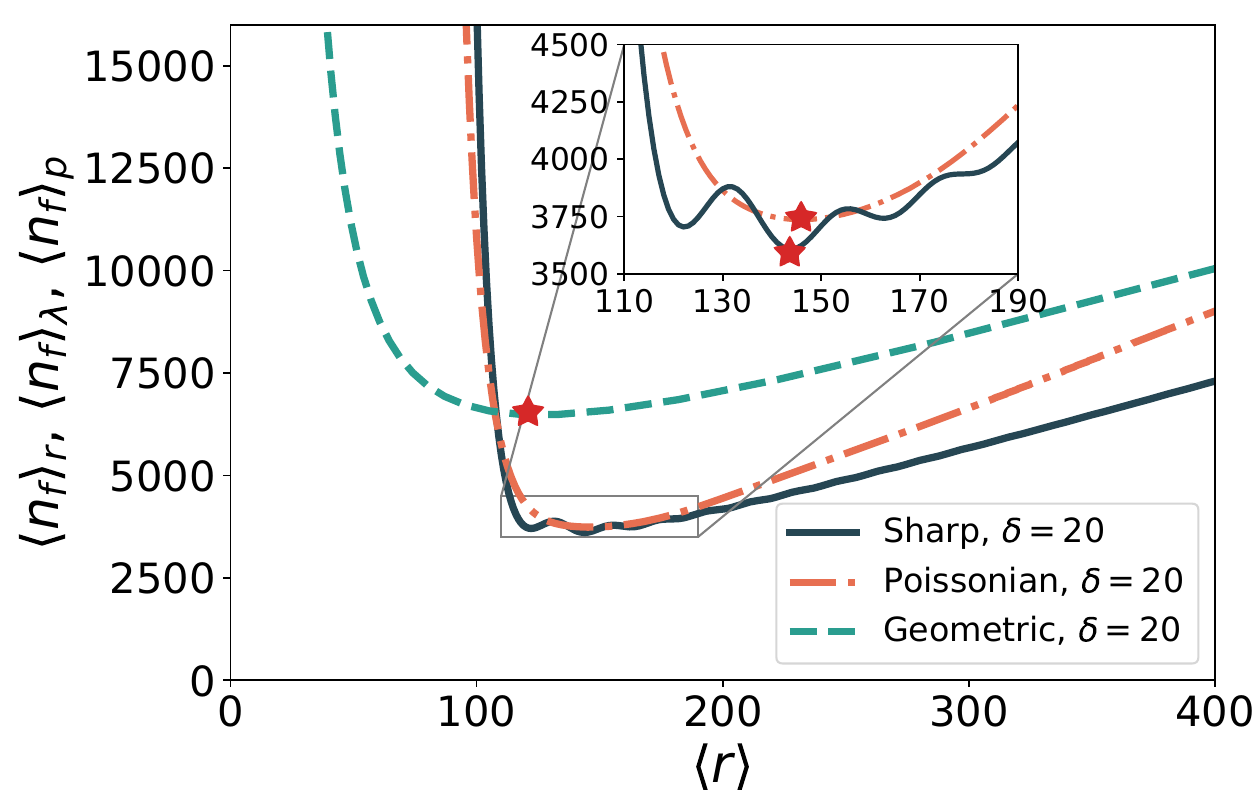}\\
\includegraphics[width=0.88\linewidth]{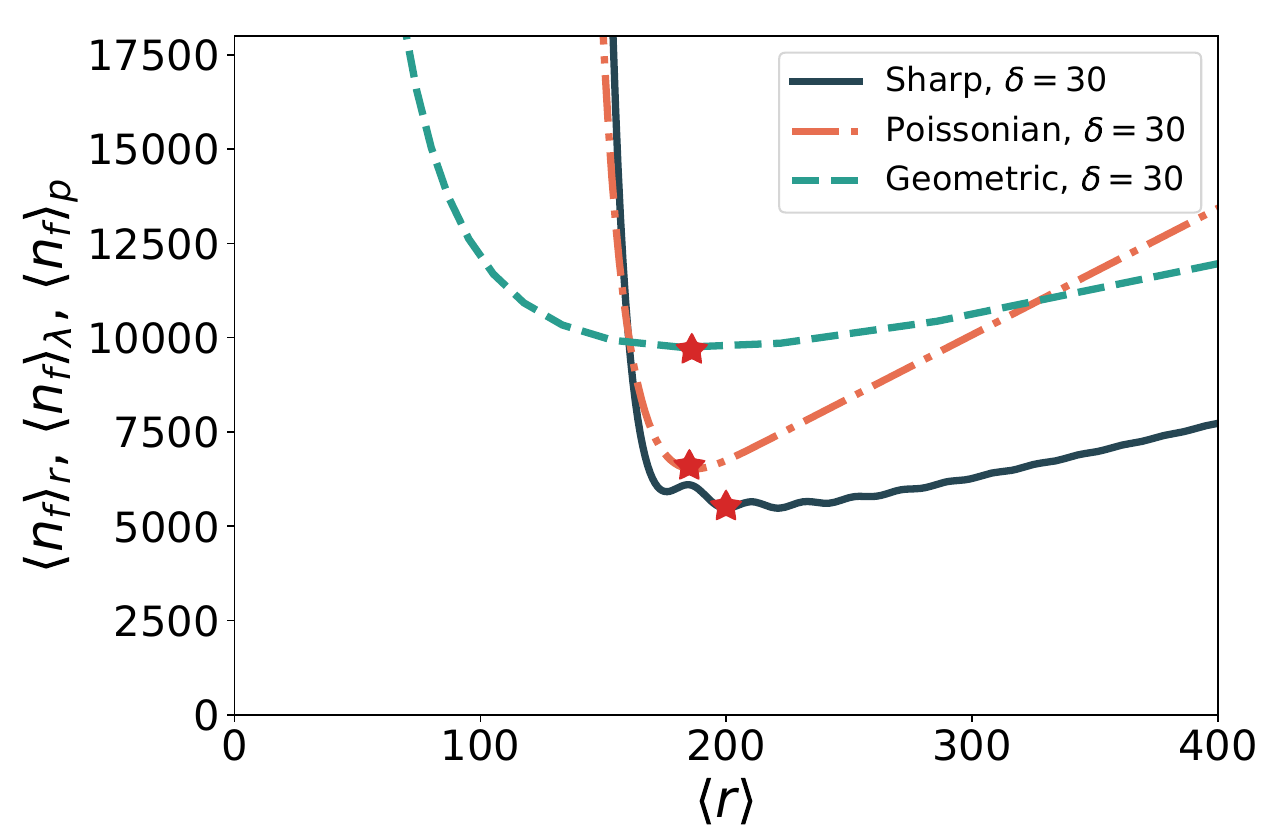}\\
\caption{A comparison between sharp, Poisson, geometric distributed restart strategies,
the mean $\bracket{n_f}_R$ vs. the mean restart step $\bracket{r}$. 
The sampling time is set as $\tau=0.1$.
As expected, the quantum oscillations are captured only by the sharp restart,
see the inset for zoom-in, and Eqs. (\ref{eqsm16b},\ref{eqsm23},\ref{eqsm25}).
As mentioned in the main text, the mean hitting time with sharp restart 
takes the lowest minimum, see the red stars denoting each minimum.
For $\delta=10$, 
$\min(\bracket{n_f}_r)=1734$, $\min(\bracket{n_f}_\lambda)=1874$, 
$\min(\bracket{n_f}_p)=3209$.
For $\delta=20$, 
$\min(\bracket{n_f}_r)=3607$, $\min(\bracket{n_f}_\lambda)=3736$, 
$\min(\bracket{n_f}_p)=6488$.
For $\delta=30$, 
$\min(\bracket{n_f}_r)=5473$, $\min(\bracket{n_f}_\lambda)=6515$, 
$\min(\bracket{n_f}_p)=9743$.
See also the inset for confirmation.
}
\label{figSM5}
\end{figure}
In the Letter we studied the quantum first detection problem with a sharp restart strategy.
We now develop the theory for more general restart processes, 
namely the case when the restart times are randomly distributed. 
More precisely, the time intervals between restarts 
are independent and identically distributed discrete random variables,
denoted with $r$. 
At the end of the section we will present the mean first detection time under restart, 
for sharp restart, Poisson restart and geometric restart, see Fig. \ref{figSM5} below. 
The goal is to study the quantum oscillations in Fig. 3 in the main text, 
and to see whether they are found also for other distributions of waiting times.
For that we need a new formalism.
We start from the general framework 
proposed recently in Refs. \cite{restart,Pal2021},
which states that the mean hitting time with a restart strategy $R$ 
(performing restarts at random time $r$) is 
%
\begin{equation}\label{eqsm15}
    \bracket{n_f}_R = {\bracket{\text{min} (n, r)} 
                      \over
                      P(n\le r)
                      }
                      ,
\end{equation}
where $n$ is the first hitting time in the absence of restart, 
and the numerator means the expectation of the minimum 
of $n$ and the random restart time/step $r$.
And hereinafter we denote $P(cond)$ 
as the probability that the condition $cond$ is satisfied.
We note that
\begin{equation}\label{eqsm15a}
\begin{aligned}
    \bracket{\text{min} (n, r)} 
    =&  \sum_{k=1}^\infty k P(r=k) 
        \left( 1-\sum_{n=1}^{k-1} F_n \right) \\
        &+
        \sum_{n=1}^\infty n F_n \sum_{k=n+1}^\infty P(r=k), \\
    P(n\le r) =& \sum_{n=1}^\infty F_n \sum_{k=n}^\infty P(r=k).
\end{aligned}
\end{equation}
%
Here we used the normalization of $P(r=k)$,
the later is the probability that the time interval between restarts is $k$. 
In what follows in the section, we will provide formulae for the mean hitting time 
with Poisson, geometric and sharp restart, 
which were already studied in \cite{Pal2021b} 
where the classical first passage time is considered.

\textbf{\em Poisson restart.}
Assuming the restart obeys the Poisson distribution,
namely,
\begin{equation}\label{eqsm16a}
    P(r=k) = {e^{-\lambda} \lambda^{k-1} \over (k-1)!}, \quad k\ge 1.   
\end{equation}
The value of $k$ is shifted so that $P(r=k)$ for $k\ge 1$ is normalized.
The mean of restart step is determined by the parameter $\lambda$, indicated by
$\bracket{r}=\sum_{k=1}^\infty k P(r=k) = 1+\lambda$.
Eq. (\ref{eqsm16a}) allows the simplification of the following:
\begin{equation}
\begin{aligned}
     &\bracket{\text{min} (n, r)} \\ 
    =&  \sum_{k=1}^\infty k {e^{-\lambda} \lambda^{k-1} \over (k-1)!} - 
        \sum_{k=1}^\infty k {e^{-\lambda} \lambda^{k-1} \over (k-1)!} 
        \sum_{n=1}^{k-1}F_n \\
        &+ \sum_{n=1}^\infty n F_n 
        \sum_{k=n+1}^\infty {e^{-\lambda} \lambda^{k-1} \over (k-1)!} \\ 
    =&  {1+\lambda} - \sum_{n=1}^\infty F_n 
        \sum_{k=n+1}^\infty k {e^{-\lambda} \lambda^{k-1} \over (k-1)!} \\ 
        &+ \sum_{n=1}^\infty nF_n 
        \sum_{k=n+1}^\infty {e^{-\lambda} \lambda^{k-1} \over (k-1)!} \\ 
    =&  {1+\lambda} - \sum_{n=1}^\infty F_n 
        \sum_{k=n+1}^\infty (k-n) {e^{-\lambda} \lambda^{k-1} \over (k-1)!}, \\  
      ~\\
     &P(n \le r) 
     =  \sum_{n=1}^\infty F_n 
        \sum_{k=n}^\infty {e^{-\lambda} \lambda^{k-1} \over (k-1)!} .
\end{aligned}
\end{equation}
We note
\begin{equation}
\begin{aligned}
     \sum_{k=n}^\infty {e^{-\lambda} \lambda^{k-1} \over (k-1)!}
    =&  1-\frac{\Gamma (n-1,\lambda )}{\Gamma (n-1)} =: \alpha(n,\lambda), \\ 
     \sum_{k=n+1}^\infty k{e^{-\lambda} \lambda^{k-1} \over (k-1)!}
    =&  \frac{e^{-\lambda } \lambda ^{n+1} - \lambda \Gamma (n+1,\lambda )}{\Gamma (n+1)}\\
        &+ \frac{e^{-\lambda}\lambda^n -\Gamma(n,\lambda) }{\Gamma (n)}+ 1+\lambda \\ 
    =:& \,\beta(n+1,\lambda).
\end{aligned}
\end{equation}
where $\Gamma(a,z)$ is the upper incomplete gamma function, 
and $\Gamma(z)$ is the gamma function.
Thus the mean hitting time with Poisson restart is
\begin{equation}\label{eqsm16b}
    \bracket{n_f}_\lambda = 
    {1+\lambda-\sum_{n=1}^\infty F_n 
    \Big[ \beta(n+1,\lambda)-n\alpha(n+1,\lambda) \Big] 
    \over
    \sum_{n=1}^\infty F_n \alpha(n,\lambda) 
    }
    .
\end{equation}
We note that in the extreme case of $\lambda=0$,
$P(r=k) = {e^{-\lambda} \lambda^{k-1} \over (k-1)!}=\delta_{k,1}$,
namely a deterministic/sharp restart at $r=1$,
and straightforward calculation also yields $\bracket{n_f}_{\lambda=0} = 1/F_1$, 
which is exactly the mean under sharp restart at $r=1$.

\textbf{\em Geometric restart.}
Considering the restart step $r$ obeying the geometric distribution,
which is a discrete version of the exponential distribution,
namely,
\begin{equation}\label{eqsm16}
    P(r=k) = (1-p)^{k-1} p, \quad k\ge 1.   
\end{equation}
We shift the conventional form by unity since the restart happens from $r=1$.
The parameter $p$ dominates the mean of restart step, 
i.e. $\bracket{r}=1/p$.
With Eq. (\ref{eqsm16}), and Eqs. (\ref{eqsm15},\ref{eqsm15a}),
we have \cite{Pal2021}
\begin{equation}
\begin{aligned}
     &\bracket{\text{min} (n, r)} \\ 
    =&  \sum_{k=1}^\infty k p(1-p)^{k-1} - 
        \sum_{k=1}^\infty k p(1-p)^{k-1} \sum_{n=1}^{k-1}F_n \\
        &+ \sum_{n=1}^\infty n F_n \sum_{k=n+1}^\infty p(1-p)^{k-1} \\ 
    =&  {1\over p} - \sum_{n=1}^\infty F_n \sum_{k=n+1}^\infty kp(1-p)^{k-1}  
        + \sum_{n=1}^\infty nF_n \sum_{k=n+1}^\infty p(1-p)^{k-1} \\ 
    =&  {1\over p} - {1\over p}\sum_{n=1}^\infty F_n (1-p)^n (1+np)  
        + \sum_{n=1}^\infty nF_n (1-p)^n \\ 
    =&  {1\over p} \left[ 1- \sum_{n=1}^\infty F_n (1-p)^n \right]
\end{aligned}
\end{equation}
and
\begin{equation}
\begin{aligned}
     &P(n \le r) = \sum_{n=1}^\infty F_n \sum_{k=n}^\infty (1-p)^{k-1} p \\ 
    =& \sum_{n=1}^\infty F_n (1-p)^{n-1} 
    =  (1-p)^{-1} \sum_{n=1}^\infty F_n (1-p)^n.
\end{aligned}
\end{equation}
Hence the mean hitting time with geometric restart, denoted by $\bracket{n_f}_p$,
is expressed as
\begin{equation}\label{eqsm23}
    \bracket{n_f}_p = {1-p \over p} {1-\sum_{n=1}^\infty F_n (1-p)^n 
                                    \over
                                    \sum_{n=1}^\infty F_n (1-p)^n 
                                    }
                                    .
\end{equation}
%

\textbf{\em Sharp restart.}
The sharp-restart strategy is deterministic,
and the probability mass function of $r$ is
\begin{equation}
    P(r=k) = \delta_{k,r^\prime}, \quad k\ge 1,
\end{equation}
which gives the mean $\bracket{r}=r^\prime$ directly.
For simplicity, we will use $r$ as the mean, to replace $r^\prime$.
Then Eq. (\ref{eqsm15}) gives
\begin{equation}\label{eqsm25}
    \bracket{n_f}_r = r{1- P_\text{det}(r)\over P_\text{det}(r)} 
                      + {\sum_{n=1}^r nF_n \over P_\text{det}(r)}.
\end{equation}
Multiplying both sides of Eq. (\ref{eqsm25}) by the measurement period $\tau$ 
gives the Eq. (3) in the main text.
Alternatively, 
we provide here another straightforward derivation for the mean in this case, 
using the first hitting probability under sharp restart.
With some fixed restart step at $r$, 
the first hitting at $n=r{\cal R}+\tilde{n}$ 
(${\cal R}\ge 0$ is the number of restarts happened, $1\le \tilde{n}\le r$)
occurs with probability
\begin{equation}
    F_n^{(r)} = \Big[ 1-P_\text{det}(r) \Big]^{\cal R} F_{\tilde{n}},
\end{equation}
where the superscript $(r)$ means the deterministic restart time at step $r$.
By definition, ${\cal R}=[(n-1)/r]$ with $[]$ means the integer part,
and $\tilde{n}=n-r{\cal R}=1+\text{mod}(n-1,r)$.
Then the mean hitting time with sharp restart is 
\begin{equation}\label{eqsm27}
\begin{aligned}
    &\bracket{n_f}_r = \sum_{n=1}^\infty n F_n^{(r)} 
    =   \sum_{r{\cal R}+\tilde{n}=1}^\infty n \left[ 1- P_\text{det}(r) \right]^{\cal R} 
        F_{\tilde{n}} \\
    =&  \sum_{{\cal R}=0}^\infty \sum_{\tilde{n}=1}^r
        \left( r{\cal R} + \tilde{n} \right)
        \left[ 1- P_\text{det}(r) \right]^{\cal R} P_\text{det}(r) 
        \,\,F_{\tilde{n}}/P_\text{det}(r) \\
    =&  r\underbrace{\sum_{{\cal R}=0}^\infty
        {\cal R}
        P_\text{det}(r) \left[ 1- P_\text{det}(r) \right]^{\cal R}}
        _{\bracket{\cal R}}
        \,  
        \underbrace{
        \sum_{\tilde{n}=1}^r {F_{\tilde{n}} / P_\text{det}(r)}
        }_{=1} \\
        &+
        \underbrace{\sum_{\tilde{n}=1}^r \tilde{n} F_{\tilde{n}} / P_\text{det}(r)}
        _{\bracket{\tilde{n}}}
        \,
        \underbrace{
        \sum_{{\cal R}=0}^\infty
        P_\text{det}(r) \left[ 1- P_\text{det}(r) \right]^{\cal R}
        }_{=1} \\
    =&  \underbrace{{r \left[ 1- P_\text{det}(r) \right] \over P_\text{det}(r)} 
        \vphantom{\sum_{\tilde{n}=1}^r n} }
        _{r\bracket{\cal R}}  +
        \underbrace{\sum_{\tilde{n}=1}^r {\tilde{n} F_{\tilde{n}} \over P_\text{det}(r)}}
        _{\bracket{\tilde{n}}}.
\end{aligned}
\end{equation}
This agrees with Eq. (\ref{eqsm25}).
The formula depicts a clear probabilistic picture:
the expectation of first hitting at $n=r{\cal R}+\tilde{n}$
is equal to the expected number of restarts happened, $\bracket{{\cal R}}$,
with ${\cal R}$ obeying the geometric distribution 
$G({\cal R}=k)=P_{\rm det}(r) [1-P_{\rm det}(r)]^k$ with $k=0,1,2,\cdots$,
multiplied by $r$,
plus the expected number of attempts till hitting following the last restart,
conditioned successful click before the next restart,
namely $\bracket{\tilde{n}}$, with $\tilde{n}$ distributed by $F_{\tilde{n}}/P_{\rm det}(r)$.

We are now ready to study the effect of random time intervals between restarts. 
In Fig. \ref{figSM5} of this supplemental material
we plot the mean of the restarted hitting time $\langle n_f \rangle_R$ 
versus the mean of restart time $r$ for the three distributions studied here.  
We do so for different initial states, 
namely the distance between the detected state 
and the initial state denoted $\delta$, 
on the one dimensional lattice is varied.  
Then we manipulate $\langle r \rangle$ (the $x$ axis of the plot), 
e.g. for Poisson case we vary the parameter $\lambda$, etc. 
Using $F_n$'s obtained from the quantum renewal equation, see Sec. \ref{secC} below, 
we can plot these curves without effort. 
The main observation are that: 
a) Poisson and geometric distributions do not give any visible oscillations 
and b) that the global minimum of the restarted process is always found 
for the sharp restart. 
The observation a) is related to the wipe out of coherence 
due to the randomness of the time interval between restarts. 
The second effect, as mentioned, is due to the general theorem studied in \cite{Luby1993},
and in the next section. 
It is left for future studies to study if other distributions of times between restarts 
can yield oscillations beyond the sharp case, 
for example very narrow distributions around the mean, 
which are not totally sharp, are expected to give some oscillations as well.

\section{Proof for the dominance of sharp-restart strategy}
Here we prove that the sharp-restart strategy 
outperforms any other restart strategy,
in the sense of always attaining the smallest minimum 
of the expected $n_f$ \cite{Luby1993,restart}.
We follow the work of mathematicians \cite{Luby1993}
and see alternative proof in Ref. \cite{restart}.

Assuming a sequence of waiting times between restarts, 
$\vec{S}=\{r_1,r_2,r_3,\dots\}$,
with $r_i$ being positive integers,
the restarts are performed at steps $n=r_1, \, r_1+r_2,\, r_1+r_2+r_3,\,\dots$,
and the process is stopped in the first hitting.
Then the possible values of $n_f$ will be $n_f=\sum_{i=1}^j r_i+\tilde{n}$,
where $j$ is the number of restarts happened, taking natural numbers, 
and $\tilde{n}$ is the time between the first hitting and the final restart,
$1\le\tilde{n}\le r_{j+1}$. 
Hence the probability of $n_f$ (with the time sequence $\vec{S}$) is 
$F_n^{\vec{S}} = \prod_{i=1}^{j} \left[ 1-P_\text{det}(r_i) \right] F_{\tilde{n}}$,
where $P_\text{det}(r_i) = \sum_{n=1}^{r_i} F_n$, 
and the mean of $n_f$ is 
\begin{equation}\label{eqsm10}
\begin{aligned}
  \bracket{n_f}_{\vec{S}} 
  =&  \sum_{i=1}^\infty \sum_{\tilde{n}=1}^{r_i} 
      \left( \sum_{j=1}^{i-1} r_j + \tilde{n} \right)
      \prod_{k=1}^{i-1} \Big[ 1-P_\text{det}(r_k) \Big] F_{\tilde{n}} \\
  =&  \sum_{i=1}^\infty  
      (r_1 + r_2 + \cdots + r_{i-1})
      \prod_{j=1}^{i-1} \Big[ 1-P_\text{det}(r_j) \Big] P_\text{det}(r_i) \\ 
  &+  \sum_{i=1}^\infty 
      \left\{
      \prod_{j=1}^{i-1} \Big[ 1-P_\text{det}(r_j) \Big] 
      \right\}
      \sum_{\tilde{n}=1}^{r_i} 
      \tilde{n} F_{\tilde{n}} \\ 
  =&  \sum_{i=1}^\infty
      r_i 
      \sum_{j=i}^\infty \prod_{k=1}^{j}
      \Big[ 1-P_\text{det}(r_k) \Big] P_\text{det}(r_{j+1}) \\ 
  &+  \sum_{i=1}^\infty 
      \left\{
      \prod_{j=1}^{i-1} \Big[ 1-P_\text{det}(r_j) \Big] 
      \right\}
      \sum_{\tilde{n}=1}^{r_i} 
      \tilde{n} F_{\tilde{n}} .
\end{aligned}
\end{equation}
%
Note that 
\begin{equation}\label{eqSM16c}
\begin{aligned}
  &\sum_{j=i}^\infty \prod_{k=1}^{j}\left[ 1-P_\text{det}(r_k) \right] 
  P_\text{det}(r_{j+1}) \\
=& \sum_{j=i}^\infty P \left( \sum_{k=1}^j r_k < n_f \le \sum_{k=1}^{j+1} r_k \right) \\ 
=& P \left( n_f > \sum_{k=1}^{i} r_k \right).
\end{aligned}
\end{equation}
We now related between $\bracket{n_f}_{\vec{S}}$ 
for a general sequence and $\bracket{n_f}_r$ for sharp restart.
Recall that for sharp restart and hence for $\bracket{n_f}_r$ 
the waiting times between restarts just become  
$\{r,r,r,\dots\}$.
We use Eq. (\ref{eqsm25}) to solve for 
$\sum_{\tilde{n}=1}^{r_i}\tilde{n} F_{\tilde{n}}$,
namely for sharp restart,
\begin{equation}\label{eqSM17}
\begin{aligned}
  &\bracket{n_f}_r = 
  r { 1-P_{\rm det}(r) \over P_{\rm det} (r)} 
  +
  \sum_{\tilde{n}=1}^r {\tilde{n} F_{\tilde{n}} \over P_{\rm det} (r)} \\
  \Rightarrow& 
  \sum_{\tilde{n}=1}^{r_i}\tilde{n} F_{\tilde{n}}
  = P_\text{det}(r_i)\bracket{n_f}_{r_i}-r_i 
    \Big[ 1-P_\text{det}(r_i) \Big],
\end{aligned}
\end{equation}
where $\bracket{n_f}_{r_i}$ is the mean of $n_f$ with sharp restart at $r_i$,
or with the waiting time (between restarts) sequence $\{r_i,r_i,r_i,\cdots\}$.
With Eqs. (\ref{eqSM16c},\ref{eqSM17}), Eq. (\ref{eqsm10}) becomes
\begin{equation}
\begin{aligned}
\bracket{n_f}_{\vec{S}}
  =&  \sum_{i=1}^\infty
      r_i 
      P \left( n_f > \sum_{k=1}^{i} r_k \right) 
    + \sum_{i=1}^\infty 
      \left\{
      \prod_{j=1}^{i-1} \Big[ 1-P_\text{det}(r_j) \Big] 
      \right\} \\ 
      &\times
      \left\{ 
        P_\text{det}(r_i)\bracket{n_f}_{r_i}-r_i 
        \Big[ 1-P_\text{det}(r_i) \Big] 
      \right\} \\ 
  =&  \sum_{i=1}^\infty 
      \left\{
      \prod_{j=1}^{i-1} \Big[ 1-P_\text{det}(r_j) \Big] 
      \right\}
      P_\text{det}(r_i)\bracket{n_f}_{r_i} \\  
  =&  \sum_{i=1}^\infty P 
      \left(\sum_{k=1}^{i-1} r_{k} < n_f \le \sum_{k=1}^{i} r_k \right) 
      \bracket{n_f}_{r_i},
\end{aligned}
\end{equation}
%
%
%
where we employed the relation 
\begin{equation}
P \left( n_f > \sum_{k=1}^{i} r_k \right) =
\prod_{j=1}^{i-1} \left[ 1-P_\text{det}(r_j)\right] \left[ 1-P_\text{det}(r_i) \right].
\end{equation}
%
%
%
%
We now use the technique presented in \cite{Luby1993}.
Since $P \left(\sum_{k=1}^{i-1} r_{k} < n_f \le \sum_{k=1}^{i} r_k \right)$ is normalized,
we obtain
%
\begin{equation}\label{eqsm14}
    \bracket{n_f}_{\vec{S}} = \sum_{i=1}^\infty \lambda_i \bracket{n_f}_{r_i} 
    \ge \sum_{i=1}^\infty \lambda_i \min(\bracket{n_f}_{r_i}) = \min(\bracket{n_f}_{r_i}),
\end{equation}
where $\lambda_i = P \left(\sum_{k=1}^{i-1} r_{k} < n_f \le \sum_{k=1}^{i} r_k \right)$
and $\sum_{i=1}^\infty \lambda_i =1$ is used.
Hence the mean of $n_f$ with the waiting time (between restarts) sequence $\vec{S}$ 
can be expressed as a convex combination of 
$\{\bracket{n_f}_{r_1}, \bracket{n_f}_{r_2}, \bracket{n_f}_{r_3}, \cdots\}$.
Therefore, 
among general restart strategies,
the mean hitting time with sharp restart 
always attains the lowest global minimum,
as pointed out originally in Refs. \cite{Luby1993,restart}.

\section{Quantum renewal equation}\label{secC}
Here we recap the first detected passage or first hitting statistics 
from the repeated-measurement protocol.
The quantum renewal equation reads \cite{Friedman2017a}
\begin{equation}\label{eqSM1}
\varphi_n 	=	\langle \delta|\hat{U}(n\tau)|0\rangle 
            - \sum_{m=1}^{n-1} 
			\langle \delta|\hat{U}[(n-m)\tau]|\delta\rangle 
            \varphi_{m}.
\end{equation}
$\varphi_n$ is the first-hitting amplitude 
whose squared absolute value gives the probability, 
namely $F_n=|\varphi_n|^2$.
The unitary evolution between measurements gives 
$\hat{U}(\tau)=\exp(-iH\tau)$
($\hbar$ is set as $1$).
Eq. (\ref{eqSM1}) shows that the first-hitting amplitude 
is equal to the measurement-free transition amplitude 
$\langle \delta|\hat{U}(n\tau)|0\rangle $ 
subtracting the measurement-free return amplitude 
$\langle \delta|\hat{U}[(n-m)\tau]|\delta\rangle$
propagating from the prior first-hitting amplitude $\varphi_m$ ($m<n$).
This is also the spirit of the classical renewal equation, 
with a replacement of the probability to amplitude \cite{Redner2001}.
One can in principal obtain all the $\varphi_n$ 
from Eq. (\ref{eqSM1}) by iterations, 
and then calculate $F_n=|\varphi_n|^{2}$, 
for all possible $\delta$ and $n$.
When $|\delta\rangle = |0\rangle$, i.e. the return case,
we present a table of the first-hitting probability $F_n$, 
for the model of an infinite line considered in the main text,
and this table is cited from Ref. \cite{Friedman2017a}, 
see Table I in the main text.
For $\delta=10$, we obtained Table \ref{tab3}.
\begin{table}[htbp]
\centering
\caption{$F_n$ for the model of an infinite line, $\delta=10$.}
\begin{tabular}{ll}
\hline \hline $\mathrm{n}$ & \multicolumn{1}{c}{$F_{n}$} \\
\hline 
1 & $| J_{10}(2 \gamma \tau )|^2$ \\
2 & $| J_0(2 \gamma \tau ) J_{10}(2 \gamma \tau )-J_{10}(4 \gamma \tau )|^2$ \\
3 & $| J_{10}(2 \gamma \tau ) J_0^2(2 \gamma \tau )-J_{10}(4 \gamma \tau ) J_0(2 \gamma \tau )$ \\ 
& $ -J_0(4 \gamma \tau ) J_{10}(2 \gamma \tau )+J_{10}(6 \gamma \tau )|^2$ \\
4 & $| J_{10}(2 \gamma \tau ) J_0^3(2 \gamma \tau )-J_{10}(4 \gamma \tau ) J_0^2(2\gamma \tau)+J_{10}(6 \gamma \tau )J_0(2 \gamma \tau)$ \\ 
& $-2 J_0(4 \gamma \tau ) J_{10}(2 \gamma \tau )J_0(2 \gamma \tau) +J_0(6 \gamma \tau ) J_{10}(2 \gamma \tau )$ \\
& $+J_0(4 \gamma \tau ) J_{10}(4 \gamma \tau )-J_{10}(8 \gamma \tau )|^2$ \\
\hline\hline
\end{tabular}
\label{tab3}
\end{table}

\section{A comparison between the results from two models}
\begin{figure}[htbp]{}
\centering
\includegraphics[width=0.9\columnwidth]{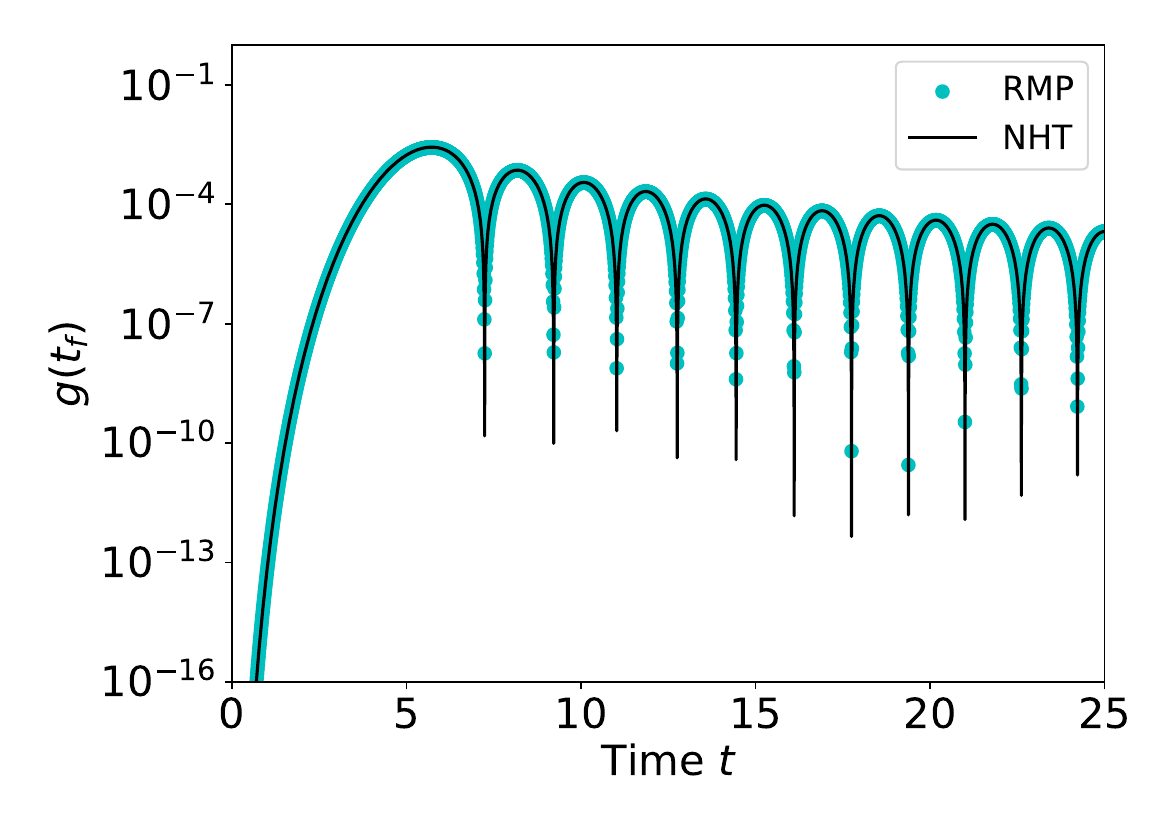}
\caption{The pdf of detection times as a function of time $t$, 
obtained from the non-Hermitian approximation Eq. (4) in the main text, 
compared with the first-hitting probability $F_n$ in the time domain.
For the latter, the data points $(n\tau, F_n/\tau)$ are plotted.
We chose $\delta=10$, and $\tau=0.05$.
}
\label{figSM1}
\end{figure}
\begin{figure}[htbp]{}
\centering
\includegraphics[width=0.9\columnwidth]{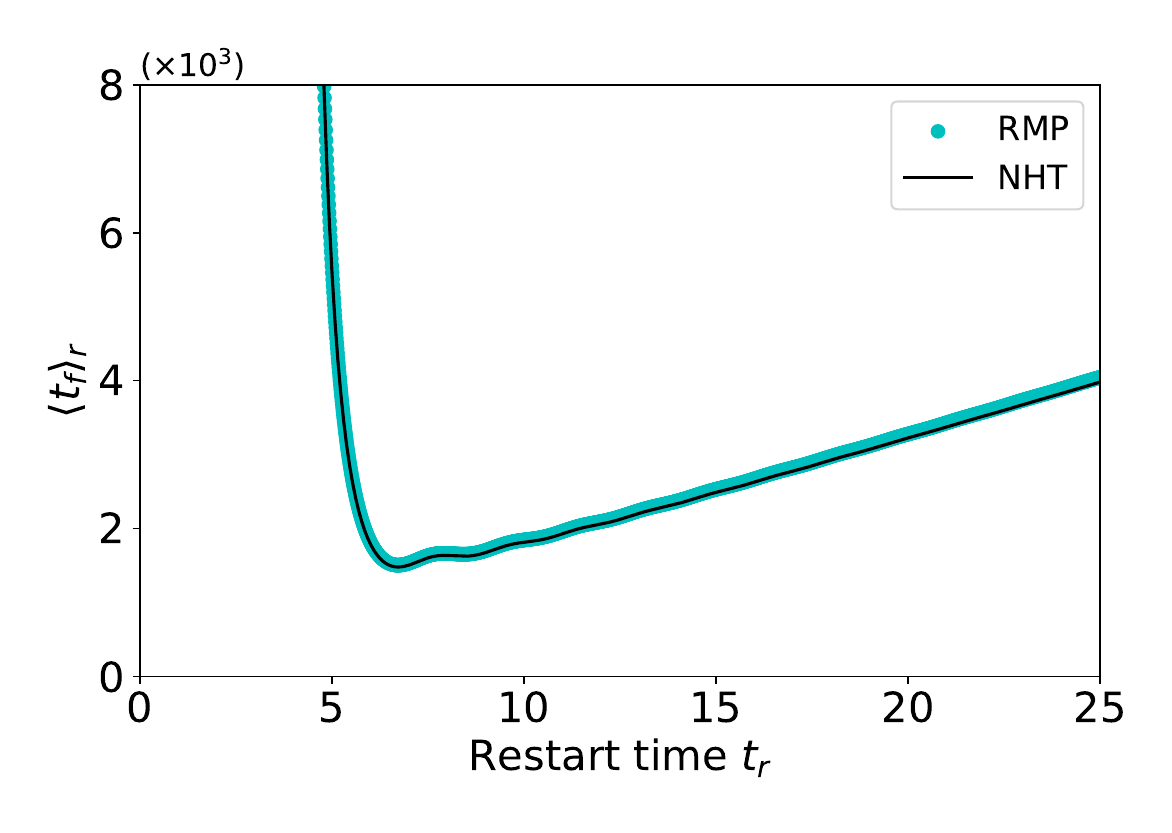}
\caption{$\langle t_f \rangle_r$ vs. restart time $t_r=r\tau$.
For $\langle n_f\rangle_r$ calculated 
from the quantum renewal equation [Eq. (\ref{eqSM1})],
the data points $(r\tau, \tau\langle n_f \rangle_r)$ are plotted.
The blue curve represents the non-Hermitian theory Eq. (5) in the main text.  
We used $\delta=10$, $\tau=0.01$.
}
\label{figSM2}
\end{figure}
Here we will compare the results obtained from the non-Hermitian theory (NHT) with those from the repeated-measurement protocol (RMP), 
which was already recapped in last section. 
Without restart, The probability density function (pdf) of detection times from NHT reads 
[Eq. (4) in the main text]
\begin{equation}\label{eqSM4}
      g(t_f) \sim \delta^2\tau {J^2_\delta(2t_f)\over t_f^2}.
  \end{equation}
And for the RMP, the pdf of detection times is actually 
$g^{rm}(t_f)=\sum_{n=1}^{\infty} F_n \delta(t_f-n\tau)$ 
where $rm$ means repeated measurements, 
with $F_n=|\varphi_n|^2$ obtained from the quantum renewal equation 
[Eq. (\ref{eqSM1})].
When plotting the RMP results for comparison, to avoid the delta function,
we will take the ``local average'' 
$(1/\tau)\int_{(n-1/2)\tau}^{(n+1/2)\tau} {\rm d}t \, g^{rm}(t)=F_n/\tau$ 
\cite{Thiel2020J}.
See Fig. \ref{figSM1} where an excellent agreement 
between the two results is shown.

For the mean detection time under restart,
Eq. (5) in the main text gives the formula derived from NHT,
and with the RMP, the mean of $n_f$ is
\begin{equation}\label{eqSM5}
\langle n_f \rangle_r = { 1-P_{\rm det}(r) \over P_{\rm det} (r)} r 
+ \langle n_f \rangle_{cond}^r,
\end{equation}
where 
$\langle n_f \rangle^r_{cond} := 
\sum_{k=1}^r k F_{k} / \sum_{j=1}^r F_{j}$.
Substituting $F_n$ into Eq. (\ref{eqSM5}) gives the mean 
$\langle n_f \rangle_r$ obtained from RMP.
The comparison between the result from Eq. (5) in the main text 
and that from Eq. (\ref{eqSM5}) evaluated numerically with the $F_n$ 
found from the renewal equation,
is presented in Fig. \ref{figSM2}. 
Clear the agreement is excellent for the small value $\tau=0.01$ under study.

\section{An explicit expression of Eq. (5) in the main text}
As shown in Eq. (5) in the main text, 
the integrals contain Bessel functions of the first kind $J_\delta(2t_f)$
and powers of $t_f$.
Further simplification gives
  \begin{equation}\label{eqSM6a}
      \bracket{t_f}_r  \sim  \left[
                          (\delta^2 \tau {\cal I}_1)^{-1} -1 
                          \right] t_r 
                        + {\cal I}_2/{\cal I}_1 , 
  \end{equation}
with
  \begin{equation}\label{eqSM6b}
  \begin{aligned}
      {\cal I}_1  &=  \frac{1}{\left(4 \delta^2-1\right) t_r}
                      [
                      \left( 8 t_r^2+2\delta+ 1 \right) 
                      J^2_\delta(2 t_r) 
                      + 8 t_r^2 J^2_{\delta+1}(2 t_r) \\
                      &\quad\quad\quad\quad\quad\quad\quad
                      - 4 t_r (2 \delta + 1) J_{\delta+1}(2 t_r) J_\delta(2 t_r)
                      ] , \\
      {\cal I}_2  &=  t_r^{2 \delta}\, \Gamma (2 \delta) \,\, 
                      _2\tilde{F}_3
                      \left(
                      \delta,\delta+1/2;
                      \delta+1,\delta+1,2\delta+1;
                      -4 t_r^2
                      \right) ,
  \end{aligned}
  \end{equation}
where $\Gamma(x)$ is the gamma function, 
and $_p\tilde{F}_q(a_1,\dots,a_p;b_1,\dots,b_q;z)$ 
is the regularized hypergeometric function, i.e.
  \begin{equation}
  \begin{aligned}
      _2\tilde{F}_3 \left(
                          \delta,\delta+1/2;
                          \delta+1,\delta+1,2\delta+1;
                          -4 t_r^2
                    \right) \\
                    =   {{}_2F_3  \left(
                                  \delta,\delta+1/2;
                                  \delta+1,\delta+1,2\delta+1;
                                  -4 t_r^2
                                  \right)
                        \over
                        \Gamma^2(\delta+1) \Gamma(2\delta+1)
                        }
                        ,
  \end{aligned}
  \end{equation}
where $_pF_q(a_1,\dots,a_p;b_1,\dots,b_q;z)$ is the hypergeometric function.
With those results, 
the limits of $\bracket{t_f}_r$ in the main text are yielded.


\section{Large $\delta$ case of Eq. (7) in the main text}
Since $F_n$ exhibits monotonical growth for small $n$ 
and decays with superimposed oscillations when $n>\delta/2\tau$
\cite{Thiel2018a},
the several minima of $\bracket{t_f}_r$ due to the oscillations 
appear when $t_r>\delta/2$.
Hence in large $\delta$ limit, the solutions to Eq. (7) in the main text,
$t_r^{ext}>\delta/2$ will be large.
With the large argument approximation for $J_\nu(x)$ and Eq. (\ref{eqSM6b}),
Eq. (7) in the main text yields
\begin{equation}\label{boundt}
{1\over2}[1+(-1)^\delta] \sin(4t_r^{ext}) =
2 (t_r^{ext}/\delta)^2 +(-1)^\delta (t_r^{ext}/\delta) \cos(4t_r^{ext}).
\end{equation}
We note that solutions to Eq. (\ref{boundt}) do not agree well 
with those to Eq. (7) in the main text,
but the aim here is to provide an upper-bound for $t_r^{ext}$, 
which is feasible by analyzing Eq. (\ref{boundt}).
The left-hand side of Eq. (\ref{boundt}) is bounded in $[0,1]$,
while the lower-bound of the right-hand side is 
$2x^{2}-x$ with $x=t_r^{ext}/\delta$.
Thus the upper-bound of $t_r^{ext}$ is given by $2x^2-x=1$,
leading to $t_r^{ext}<\delta$.
Namely, the minima of $\bracket{t_f}_r$ appear 
on the interval $t_r\in[\delta/2, \delta]$.

\subsection{Details for $t_r$ vs. $\delta$}
\begin{figure}[ht]{}
\centering
\includegraphics[width=0.9\columnwidth]{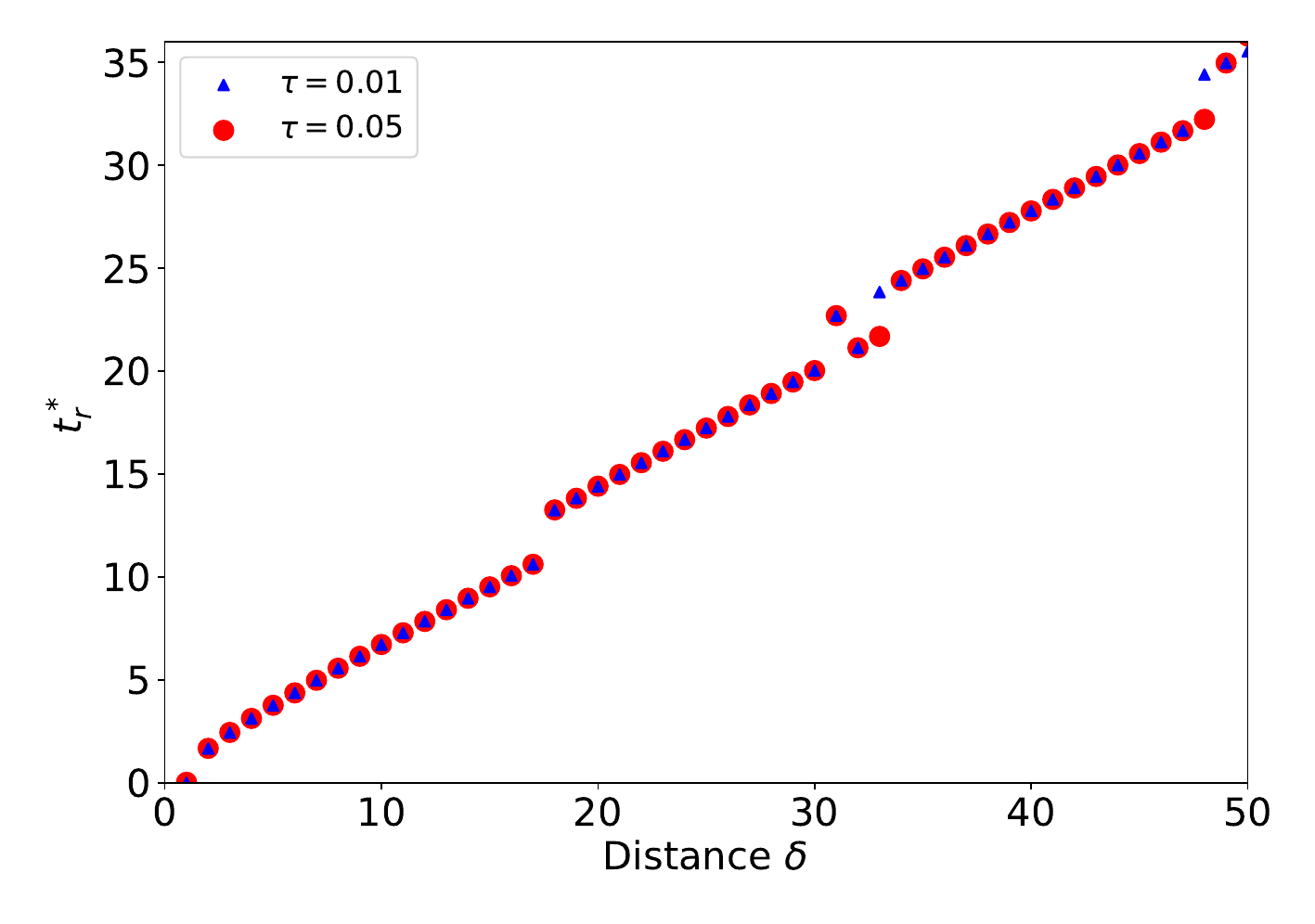}
\caption{The optimal restart time $t_r$ versus $\delta$ for $\tau=0.01, 0.05$,
obtained from the non-Hermitian theory.
As shown, $t_r$ increases roughly linear in $\delta$ 
due to the ballistic spreading of the quantum wave packet,
and exhibits sudden jumps at some special $\delta$'s,
which is speculated as being homologous to the plunges in large $\tau$ limit.
}
\label{figSM2a}
\end{figure}

\begin{figure}[ht]{}
\centering
\includegraphics[width=0.9\columnwidth]{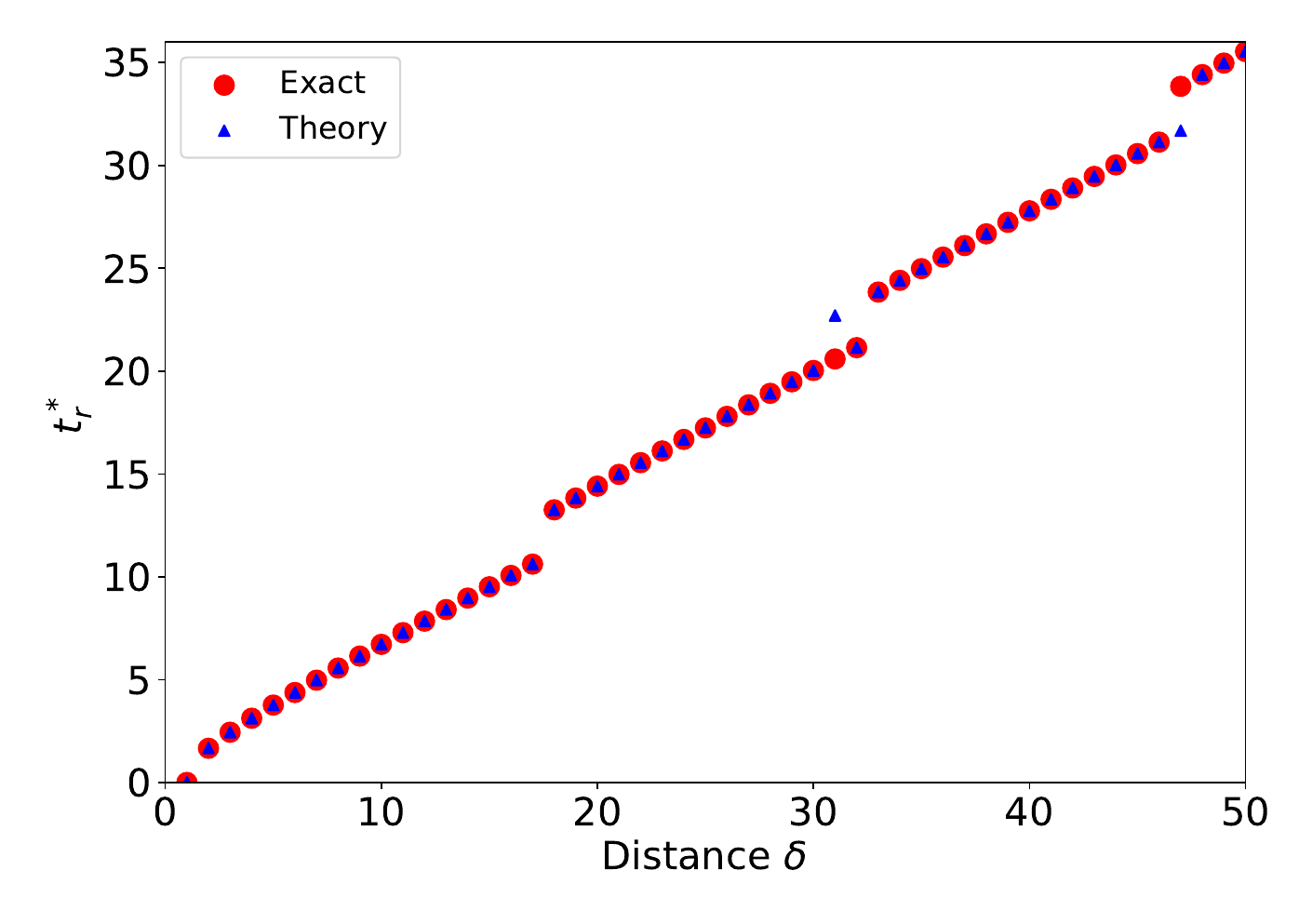}
\caption{The optimal restart time $t_r$ versus $\delta$ for $\tau=0.01$.
We compare the theoretical and exact results. 
The theory agrees with the exact results nicely, except on the blips. 
}
\label{figSM2b}
\end{figure}

We present here the theoretical optimal restart time
$t_{r}$ as a function of $\delta$ for $\tau=0.01$ and $\tau=0.05$,
and also compare the theory with the exact results.
The theory is based on the minimization of $\bracket{t_f}_r$ 
from Eq. (5) in the main text,
and the exact results are obtained from the the repeated-measurement model,
namely minimizing $\bracket{n_f}_r$ calculated with $F_n$.
For the latter, we used $t_r^*=r^*\tau$.
Fig. \ref{figSM2a} shows the $\tau$-independence of $t_{r}$,
as well as the non-trivial behavior of sudden jumps (as mentioned in the main text), 
which might be homologous to the plunges in large $\tau$ limit.
Fig. \ref{figSM2b} shows the comparison between the theory and exact results for $\tau=0.01$, 
and the case $\tau=0.05$ was also check by the authors without showing here.

\section{The mean first hitting time under sharp restart for the classical model}
\begin{figure}[htbp]{}
\centering
\includegraphics[width=0.95\linewidth]{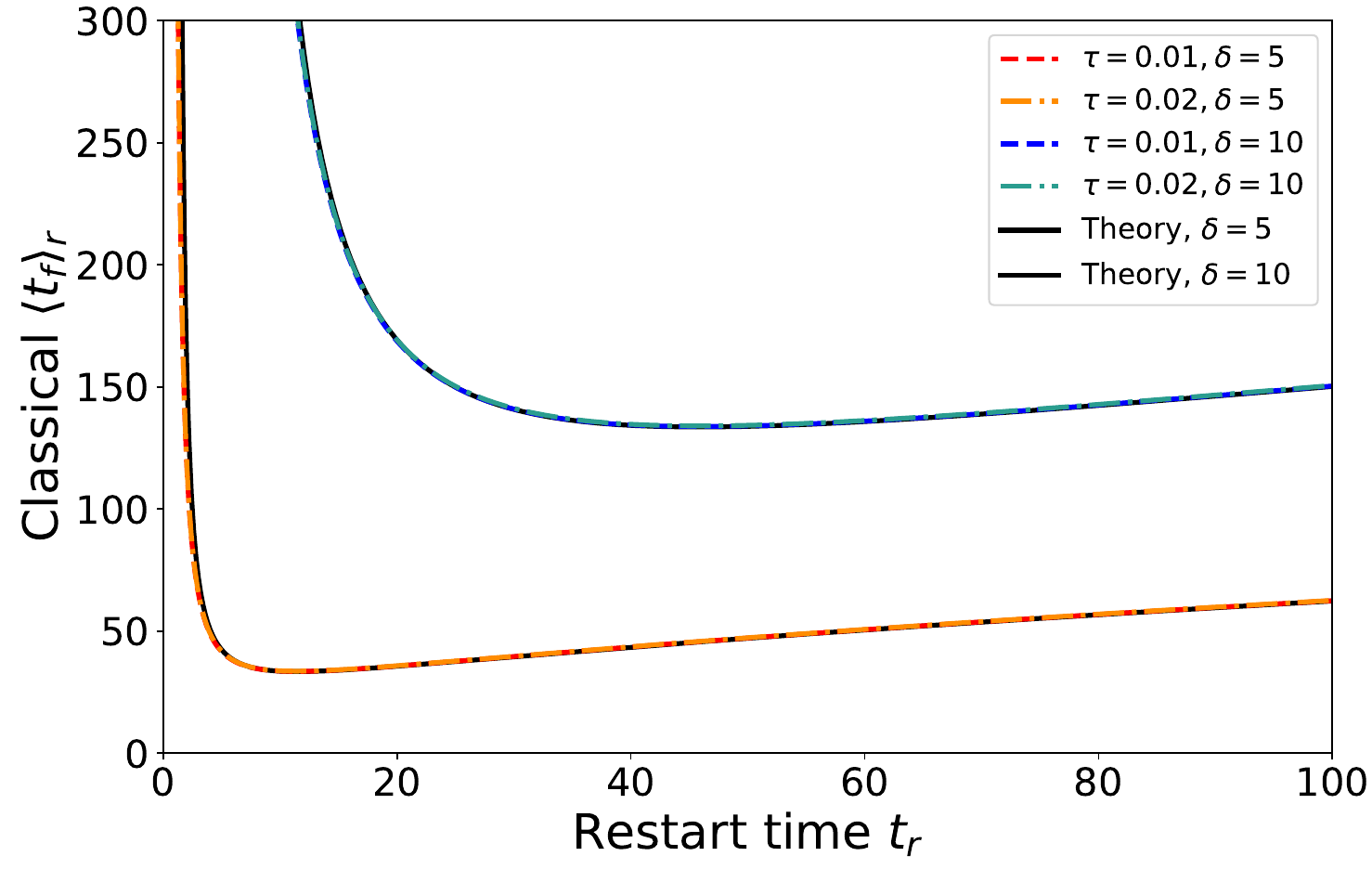}
\caption{$\bracket{t_f}_r$ vs. $t_r$ for different $\delta$ and $\tau$.
The model here is a classical random walk in one dimension 
(as mentioned in the Introduction in the main text),
and we employ the periodical monitoring again on this model 
to make a comparison with the quantum case.
As seen, there is no counterpart of the quantum oscillations in the classical model.
The black lines represent the theoretical convergence to a diffusion process under restart 
[Eq. (\ref{cltr})].
And the colored dashed curves represent the results 
obtained from the classical renewal equation [Eq. (\ref{renewal})]
and Eq. (3) in the main text.
}
\label{figSM2c}
\end{figure}
To make a comparison between the classical and quantum restart,
we apply the sharp-restart strategy to a classical walk on the integers 
(the model mentioned in the Introduction in the main text)
under periodical monitoring. 
More specifically, the random walk is started on the origin, 
and the detection is on lattice point $\delta$. 
Every $\tau$ units of time, the experimentalist measures 
if the particle is on $\delta$ or not 
(thus we are dealing with a problem of first detection 
and not first passage like in the quantum setting). 
Restarts are sharp and take place every $r$ units of time.
To summarize the detection protocol is exactly like the one introduced in the main text 
for the quantum particle, but the dynamics between measurements is classical.

As for the quantum case studied in the Letter we focus on the small and large $\tau$ limit.
When $\tau$ is large, such that $\delta^2 / D$ is much smaller than $\tau$, 
where $D$ is the diffusion constant, i.e. $D=1$ in our case 
(with the hopping rate $\gamma$ set as $1$), 
we expect simple behavior. 
Namely that the most efficient restart is found for $r=1$. 
Simply stated, as the packet is spreading for large $\tau$, 
it is better to speed up the restarts. 
Mathematically, this is related to the observation 
that the classical first hitting probability $F_n^{\text{cl}}(\delta)$ in this limit 
is a monotonically decaying function of $n$. 
The significance of this, is that the staircases found for large $\tau$, 
in the quantum world (see Fig. 3 in the main text) 
will not be found for the classical case. 


We now focus on small $\tau$ limit of the classical restart. 
Unlike the quantum case, the classical counterpart does not suffer from the Zeno physics.
Let us focus on the general formalism of this problem, valid for any $\tau$. 
The first thing to do is to calculate $F_n^\text{cl}(\delta)$ 
the first hitting probability, in the classical domain.
Using the solution mentioned in the main text, 
that the probability of finding the classical walker on $x$ at time $t$ is
$P^{{\rm cl}}(x,t) = i^{-x} e^{-2\gamma t} J_{x}(i 2\gamma t)$,
and the classical renewal equation 
\begin{equation}\label{renewal}
    F_n^{\text{cl}}(\delta) = P^{\text{cl}}(\delta,n\tau)
                              - \sum_{m=1}^{n-1} P^{\text{cl}}[0,(n-m)\tau] 
                              F_{m}^{\text{cl}}(\delta),
\end{equation}
%
one can readily obtain the first hitting statistics.
Using those $F_n^{\text{cl}}(\delta)$, with the general formula Eq. (3) in the main text,
the mean hitting time under sharp restart and discrete sampling 
can be calculated for example numerically.
As seen in Fig. \ref{figSM2c}, for small $\tau$,
the mean hitting time under sharp restart exhibits one distinct minimum
and smooth landscape without oscillations.
Hence, the oscillations in Fig. 3 in the main text (which focus on small $\tau$ limit) 
have no counterpart in the classical model.

Let us analyze the small $\tau$ limit more carefully.
What is expected is the $\tau$-independence 
of $\langle t_f \rangle_r$ in the small $\tau$ limit,
since the dynamics of a classical random walk under discrete monitoring 
converges to a diffusion process, i.e. Brownian motion.
The classical first passage probability for a diffusive walker
is known as the Smirnof distribution \cite{Redner2001}
\begin{equation}
    f^{\rm cl} (\delta,t) = {\delta \over \sqrt{4\pi Dt^3}} e^{-\delta^2/4Dt}.
\end{equation}
As mentioned, with the hopping rate $\gamma$ in Eq. (1) in the text set as $1$,
the diffusion coefficient $D=1$.
And the continuous time limit of Eq. (3) in the main text gives
\begin{equation}\label{cltr}
\begin{aligned}
\langle t_f \rangle_r &= t_r (1-I_1)/I_1 + I_2/I_1 \\
\text{with} \quad
{I}_1 &= 
\text{erfc}\left(\frac{\delta }{2 \sqrt{t_r}}\right), \\
I_2 &=
\frac{\delta \sqrt{t_r} \, e^{-\delta^2 /4 t_r}}{\sqrt{\pi }}
-\frac{\delta^2}{2} \text{erfc}\left(\frac{\delta }{2 \sqrt{t_r}}\right),
\end{aligned}
\end{equation}
where $\text{erfc}(x)=1-\text{erf}(x)$ is the complementary error function,
with $\text{erf}(x)$ the error function.
Eq. (\ref{cltr}) gives the theoretical results shown in Fig. \ref{figSM2c},
which reach excellent agreement with the numerics calculated from Eq. (\ref{renewal}).
The analysis in this section while focusing on small and large $\tau$ limits, 
will be extended elsewhere.

\section{Derivation of Eq. (10) in the main text}
Substituting the expression of $F_n$ into Eq. (9) in the main text, 
we get
\begin{equation}\label{eqSM6}
    \sum_{n=1}^r {1\over n} \cos^2 \left( 2 n \epsilon  - {\pi\over4} \right) 
    = {r \over r+1} \cos^2 \left[ 2 (r+1) \epsilon - {\pi\over4} \right], 
\end{equation}
where $\epsilon$ is defined as in the main text, 
$\tau=k \pi/2 + \epsilon$ with $0<\epsilon<\pi/2$.
The $\epsilon$ solving Eq. (\ref{eqSM6}) gives the special $\epsilon$ 
at which the transition from $r^*=r$ to $r^*=r+1$ takes place. 
As mentioned if $\epsilon=0$, $r^*=1$. 
We denote $\epsilon_{1\to2}$ as the value of $\epsilon$ 
where we have a transition from $r^*=1$ to $r^*=2$, 
similarly for other transitions.
In between the transition $\epsilon$,
namely for each interval $[\epsilon_{k\to k+1}, \epsilon_{k+1\to k+2}]$,
we will check whether $\langle n_f \rangle_{k+1}$ remains the minimum, 
and especially compare it with $\langle n_f \rangle_1$ 
in case we miss the plunge to $r^*=1$.
We get the following:
\begin{equation}\label{eqSM7}
\begin{aligned}
    %
    &\epsilon_{1\to2} = 0.850, \\
    &\epsilon \in [\epsilon_{1\to2}, \epsilon_{2\to3}]:
    \langle n_f \rangle_2 < \langle n_f \rangle_1, \\
    &\epsilon_{2\to3} = 1.081, \\
    &\epsilon \in [\epsilon_{2\to3}, \epsilon_{3\to4}]:
    \langle n_f \rangle_3 < \langle n_f \rangle_1, \\
    &\epsilon_{3\to4} = 1.204, \\
    &\epsilon \in [\epsilon_{3\to4}, \epsilon_{4\to5}]:
    \langle n_f \rangle_4 < \langle n_f \rangle_1, \\
    &\epsilon_{4\to5} = 1.280, \\
    &\epsilon \in [\epsilon_{4\to5}, \epsilon_{5\to6}]:
    \langle n_f \rangle_5 < \langle n_f \rangle_1, \\
    &\epsilon_{5\to6} = 1.332, \\
    &\epsilon \in [\epsilon_{5\to6}, 1.353]:
    \langle n_f \rangle_6 < \langle n_f \rangle_1; \\
    &\epsilon \in [1.353, \pi/2]:
    \langle n_f \rangle_1 < \langle n_f \rangle_6. \\
\end{aligned}
\end{equation}
We note that Eq. (\ref{eqSM6}) gives more than one solutions for given $r$.
For instance, for $r=1$, the solutions are 0.850 and 1.293,
and a careful check on $\langle n_f \rangle_r$ 
excludes 1.293 since $r^*=5$ at this $\epsilon$.
Eq. (10) in the main text is obtained then.

%
\begin{figure}[ht]{}
\centering
\includegraphics[width=0.98\columnwidth]{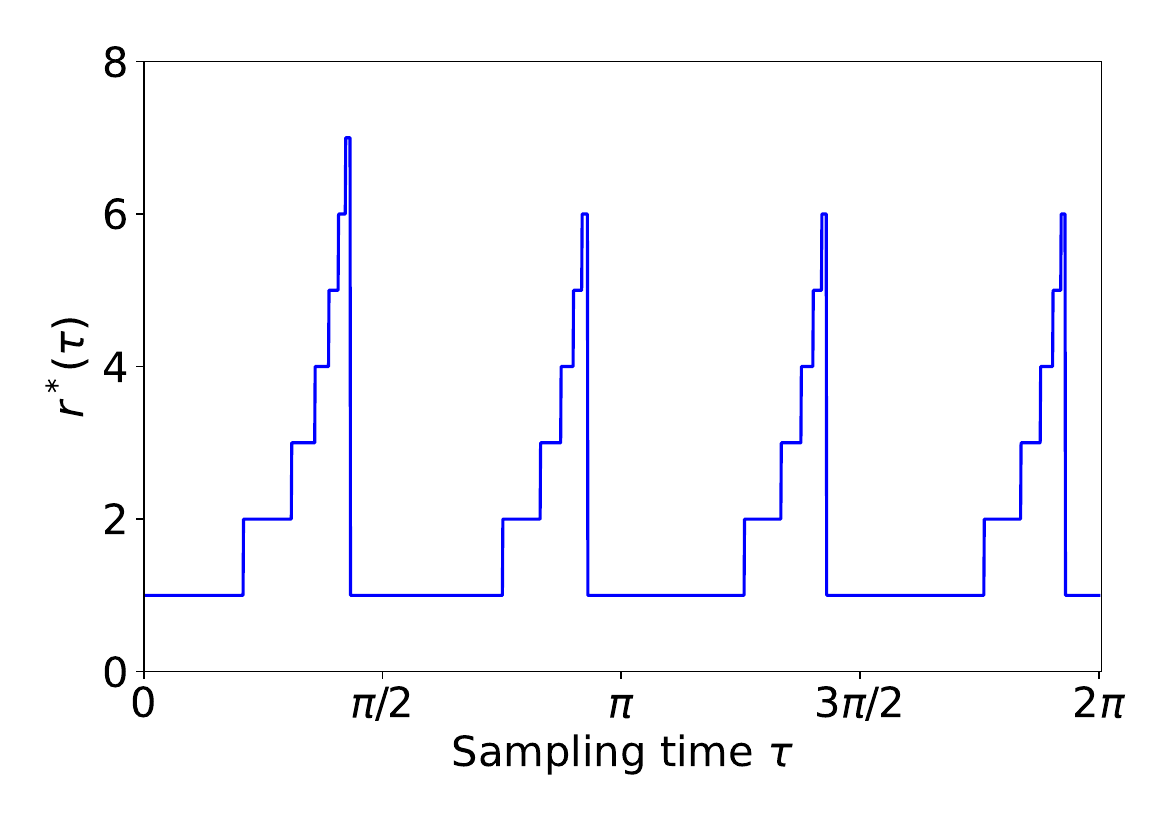}
\caption{The optimal restart time $r^*$ 
as a function of $\tau$ for $\delta=0$.
We see the staircase structure appears when $\tau$ is not large. 
}
\label{figSM3}
\end{figure}
\section{$r^*$ for the case $\delta=0$ on the full range of $\tau$}
As mentioned, for large sampling time $\tau$, 
the optimal restart step $r^*$ as a function of $\tau$
exhibits staircase structure presented in Fig. 3 in the main text.
Actually, for the return case $\delta=0$, 
the staircase structure appears on the full range of $\tau$.
We show in Fig. \ref{figSM3} the $r^*$ versus $\tau$ with $\tau\in[0,2\pi]$.
The staircase composed by $r^*=1,2,\dots,5,6$ starts from $\pi/2$, 
and before that is a staircase composed by $r^*=1,2,\dots,5,6,7$.



\begin{thebibliography}{64}
\bibitem{Remark} 
Eq. (8) is valid for not too large $n$, 
however, since our main results focus on finite $r$ restart, 
the approximation Eq. (8) holds.
\end{thebibliography}
\end{document}